\documentclass[prx,superscriptaddress,aps,amsmath,amssymb,amsfonts,floatfix,reprint,raggedbottom,longbibliography]{revtex4-2}

\usepackage{graphicx}
\graphicspath{{Figures/}}
\usepackage{balance,wrapfig}
\usepackage{dcolumn}
\usepackage{xcolor}
\usepackage{lipsum}
\usepackage{soul}
\usepackage{braket}
\usepackage[compact]{titlesec}
\usepackage{multirow}
\usepackage{enumitem}
\usepackage{array}
\usepackage{booktabs}
\usepackage[ruled,vlined,linesnumbered]{algorithm2e}
\usepackage{footnote}
\usepackage{titlesec}
\usepackage{lmodern}
\usepackage{times}
\usepackage[utf8]{inputenc}
\usepackage[subpreambles=true]{standalone}
\usepackage[para]{footmisc}
\usepackage[misc]{ifsym}
\usepackage{mathtools}
\usepackage{hyperref}

\hypersetup{
    colorlinks=true,       
    linkcolor=red,          
    citecolor=blue,        
    filecolor=magenta,      
    urlcolor=red,           
    runcolor=cyan
}

\usepackage{xr-hyper}
\usepackage{float}
\usepackage{cleveref}       

\DeclareUnicodeCharacter{2009}{\,}

\makeatletter
\newcommand*{\addFileDependency}[1]{
  \typeout{(#1)}
  \@addtofilelist{#1}
  \IfFileExists{#1}{}{\typeout{No file #1.}}
}
\makeatother

\makeatletter 
\renewcommand{\fnum@figure}{\textbf{Fig.~\thefigure}}
\makeatother

\makeatletter
\def\bbordermatrix#1{\begingroup \m@th
  \@tempdima 4.75\p@
  \setbox\z@\vbox{%
    \def\cr{\crcr\noalign{\kern2\p@\global\let\cr\endline}}%
    \ialign{$##$\hfil\kern2\p@\kern\@tempdima&\thinspace\hfil$##$\hfil
      &&\quad\hfil$##$\hfil\crcr
      \omit\strut\hfil\crcr\noalign{\kern-\baselineskip}%
      #1\crcr\omit\strut\cr}}%
  \setbox\tw@\vbox{\unvcopy\z@\global\setbox\@ne\lastbox}%
  \setbox\tw@\hbox{\unhbox\@ne\unskip\global\setbox\@ne\lastbox}%
  \setbox\tw@\hbox{$\kern\wd\@ne\kern-\@tempdima\left[\kern-\wd\@ne
    \global\setbox\@ne\vbox{\box\@ne\kern2\p@}%
    \vcenter{\kern-\ht\@ne\unvbox\z@\kern-\baselineskip}\,\right]$}%
  \null\;\vbox{\kern\ht\@ne\box\tw@}\endgroup}
\makeatother

\newcolumntype{L}{>{\arraybackslash}m{3.9 cm}}
\newcolumntype{C}{>{\centering\arraybackslash}m{3.2 cm}}
\newcolumntype{G}{>{\centering\arraybackslash}m{1.41 cm}}

\emergencystretch=\maxdimen
\titlespacing{\section}{0pt}{*3}{*2}
\titlespacing{\subsection}{0pt}{*2}{*2}
\titlespacing{\subsubsection}{0pt}{*2}{*2}

\titleformat{\section}{\filcenter\normalfont\small \bfseries}{\thesection.}{1em}{\MakeUppercase}   

\begin{document}

\title{Probabilistic Computers for MIMO Detection: From Sparsification to 2D Parallel Tempering}

\author{M Mahmudul Hasan Sajeeb}
\email{msajeeb@ece.ucsb.edu}
\affiliation{Department of Electrical and Computer Engineering, University of California, Santa Barbara, Santa Barbara, CA, 93106, USA}
\affiliation{Equally contributing authors}

\author{Kevin Callahan-Coray}
\email{kevincallahan-coray@ucsb.edu}
\affiliation{Department of Electrical and Computer Engineering, University of California, Santa Barbara, Santa Barbara, CA, 93106, USA}
\affiliation{Equally contributing authors}

\author{\mbox{Corentin Delacour}}
\affiliation{Department of Electrical and Computer Engineering, University of California, Santa Barbara, Santa Barbara, CA, 93106, USA}

\author{Sanjay Seshan}
\affiliation{Department of Electrical and Computer Engineering, Carnegie Mellon University, Pittsburgh, PA, 15213, USA}

\author{Tathagata Srimani}
\affiliation{Department of Electrical and Computer Engineering, Carnegie Mellon University, Pittsburgh, PA, 15213, USA}

\author{Kerem Y. Camsari}
\email{camsari@ece.ucsb.edu}
\affiliation{Department of Electrical and Computer Engineering, University of California, Santa Barbara, Santa Barbara, CA, 93106, USA}

\date{\today}
\begin{abstract}
Probabilistic computers built from p-bits offer a promising path for combinatorial optimization, but the dense connectivity required by real-world problems scales poorly in hardware. Here, we address this through graph sparsification with auxiliary copy variables and demonstrate two fully on-chip parallel tempering solvers on an FPGA. Targeting MIMO detection, a dense, NP-hard problem central to wireless communications, we first fit 11 temperature replicas of a 128-node sparsified system (1,408 p-bits) on-chip and achieve bit error rates significantly below conventional linear detectors on $64 \times 64$ BPSK MIMO. We report complete end-to-end solution times of 3~ms per instance, including all loading, sampling, readout, and verification overheads. ASIC projections in 7~nm technology indicate 103~MHz operation at 285.8~mW, suggesting that massive parallelism across multiple chips could approach the throughput demands of next-generation wireless systems. Sparsification, however, introduces a sharp sensitivity to the copy-constraint strength $P$ that requires manual tuning. To eliminate this bottleneck, we utilize Two-Dimensional Parallel Tempering (2D-PT), which exchanges replicas across both temperature ($\beta$) and constraint ($P$) dimensions. On Sherrington--Kirkpatrick spin glasses, 2D-PT converges roughly $250\times$ faster than optimally tuned 1D-PT, and on $128 \times 128$ MIMO it reaches zero bit errors at high SNR where 1D-PT exhibits an error floor. We further validate 2D-PT entirely on-chip with 54 replicas (1,728 p-bits) on a $16 \times 16$ MIMO instance, where it tracks the maximum-likelihood bound in just 50 Monte Carlo steps---$10\times$ fewer than 1D-PT---at projected 111~MHz and 124~mW in 7~nm. Together, these results establish an on-chip p-bit architecture and a scalable, tuning-free algorithmic framework for dense combinatorial optimization.
\end{abstract}


\maketitle

\twocolumngrid

\section{Introduction}
\label{sec:Intro}

Probabilistic computers built from p-bits have emerged as a promising architecture for combinatorial optimization and probabilistic sampling~\cite{chowdhury2025probabilistic}. By mapping hard problems to Ising Hamiltonians and leveraging the natural dynamics of coupled stochastic units, Ising machines~\cite{Mohseni2022} can explore solution spaces efficiently. However, scaling p-bit systems to real-world problem sizes faces a fundamental challenge: the dense, all-to-all connectivity required by many applications grows quadratically, causing routing congestion and limiting clock frequency in CMOS and FPGA implementations.

Here we report a fully on-chip probabilistic computer built from p-bits for dense combinatorial optimization, benchmarked on MIMO detection from wireless communications~\cite{singh2022ising, singh2024uplink}. For $64\times 64$ BPSK MIMO, the solver achieves bit error rates significantly below conventional linear detectors in 3~ms per instance. This 3~ms is the fully accounted wall clock time, averaged over 130,000 instances, with every engineering overhead included: instance preparation, weight loading, on-chip energy evaluation, readout, and verification against the original dense Hamiltonian. End-to-end accounting at this level is rare in the Ising-machine literature, yet it is what determines whether a solver can meet a system-level latency target. Two methods make it possible: graph sparsification~\cite{aadit2022massively, sajeeb2025scalable} which makes a multi-replica parallel-tempering engines routable; and on-chip Two-Dimensional Parallel Tempering (2D-PT)~\cite{delacour2025two}, which removes both the penalty-strength bottleneck of sparsified machines and the host control loop.

The challenge with hardware connectivity is not unique to p-bits. Quantum annealers~\cite{johnson2011quantum, king2023quantum}, coherent Ising machines~\cite{mcmahon2016fully,honjo_2021}, memristive arrays~\cite{fahimi_2021,jiang_2023, he2025hardware}, and coupled oscillator networks~\cite{mallick2020using,moy20221,lo2023ising,Graber_2024} all confront the same connectivity bottleneck when targeting dense problem graphs. Prior work has addressed this through minor embedding~\cite{choi2011minor,sugie2020minor} or problem decomposition, but these approaches incur significant overhead in qubit/spin count~\cite{hamerly_2019} or require iterative off-chip communication. An alternative is graph sparsification whereby auxiliary copy nodes distribute connections across a sparse physical network with bounded local degree~\cite{aadit2022massively, sajeeb2025scalable}.

\begin{figure*}[t!]
    \centering
    \includegraphics[width =1 \textwidth ]{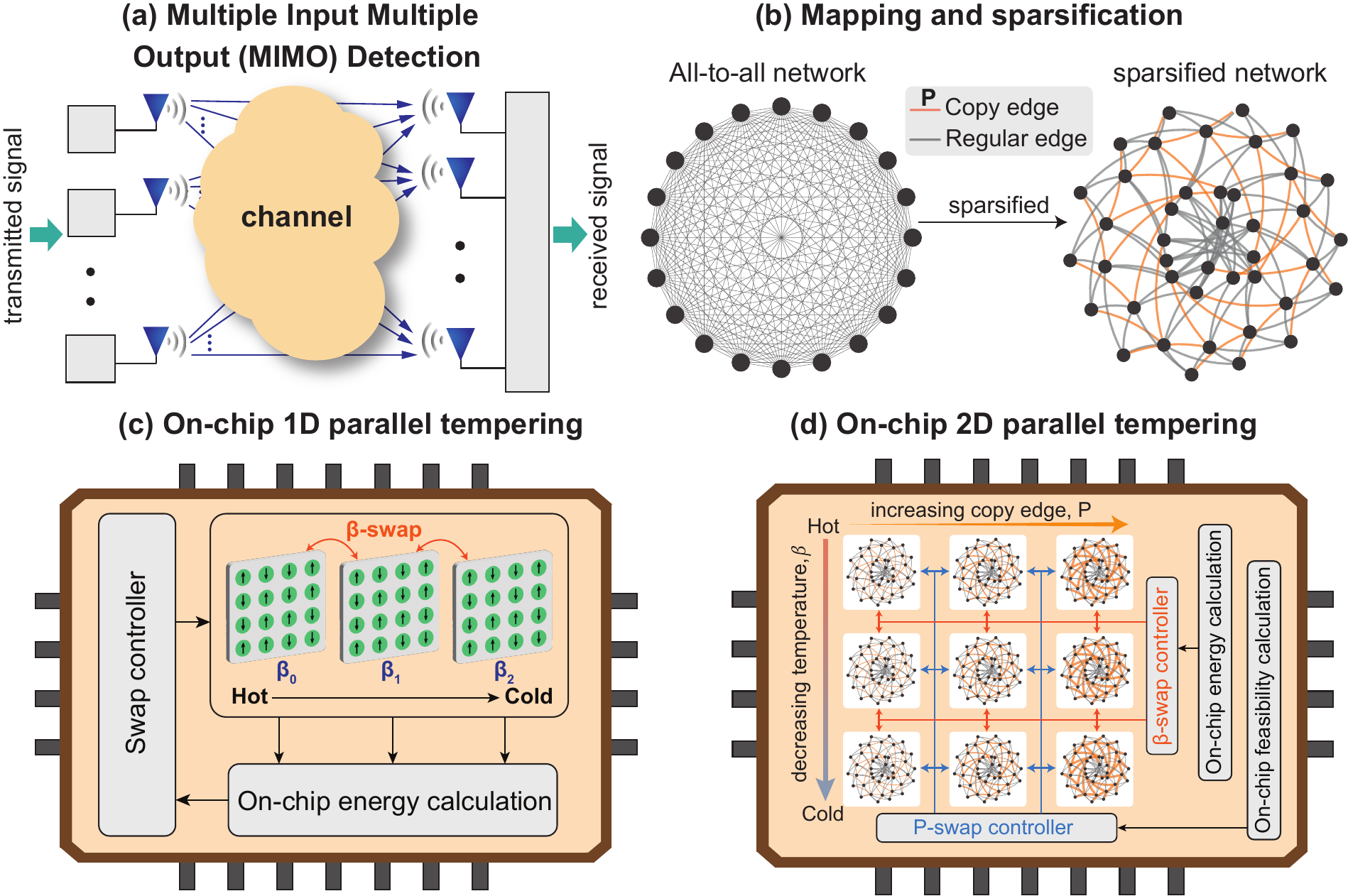}
    \caption{(a) Multiple-input multiple-output (MIMO) detection: transmitted symbols pass through a noisy fading channel and must be recovered at the receiver. (b) The maximum-likelihood detection cost maps to an Ising Hamiltonian with all-to-all connectivity (left). Sparsification introduces auxiliary copy nodes connected by ferromagnetic couplings of strength $P$ (right). (c) On-chip parallel tempering with temperature replicas instantiated entirely on an FPGA. Replicas exchange states along the inverse temperature ($\beta$) axis while $P$ remains fixed. (d) On-chip Two-Dimensional Parallel Tempering (2D-PT) arranges replicas on a grid, enabling exchanges along both $\beta$ and $P$ axes. This eliminates the need for manual $P$-tuning while improving mixing across energy barriers.}
    \label{fig:fig1}
\end{figure*}

MIMO detection exemplifies this challenge. The maximum-likelihood detector for this wireless communications problem maps to a dense Ising Hamiltonian and scales exponentially with system size~\cite{singh2022ising, singh2024uplink}, while linear approaches such as MMSE sacrifice accuracy. This complexity--accuracy tradeoff has motivated numerous chip-level~\cite{han2022low, chi2025multilinear, xiang2025area, castaneda2022283} and physics-inspired approaches for MIMO, including coherent Ising machines~\cite{singh2022ising, singh2024uplink}, oscillator networks~\cite{sreedhara2023mu, sagan2022implementing}, quantum annealers~\cite{krikidis2024mimo, kim2024x}, and probabilistic Ising machines with inertia that enable fully parallel spin updates~\cite{Roumin2026fully}. This latter work provides an interesting and complementary route to p-bit hardware scalability: it modifies the spin-update dynamics to overcome the sequential-update bottleneck, whereas the present work addresses the dense-connectivity and copy-constraint tuning bottlenecks through sparsification and two-dimensional parallel tempering. MIMO detection thus serves as an ideal benchmark for evaluating complementary routes toward scalable on-chip Ising architectures.

While sparsification solves the routing problem and enables constant-frequency scaling, it fundamentally alters the optimization landscape. It is useful to distinguish this approach from standard minor graph embedding. In minor graph embedding, the physical hardware graph is fixed in advance, as in quantum annealers whose qubit connectivity is determined by fabrication~\cite{choi2011minor,sugie2020minor}. The task is then to fit a dense logical problem graph into this pre-existing sparse topology by forming chains of physical qubits. In contrast, our sparsification framework does not target a fixed hardware graph. The dense logical graph is first rewritten into a bounded-degree sparse graph using auxiliary copy nodes, and the FPGA/ASIC architecture is then synthesized around the resulting graph~\cite{aadit2022massively,sajeeb2025scalable}. Thus, the physical design follows the sparsified problem graph, rather than the problem being forced to conform to a pre-existing hardware graph.

Despite this architectural flexibility, sparsification still introduces a new form of overhead. The logical identity between a node and its copies must be enforced through ferromagnetic couplings with strength $P$. If $P$ is too weak, copies disagree, and the solution becomes infeasible; if $P$ is too strong, the energy landscape becomes rigid, trapping the system in local minima. This critical tuning problem has limited the practical deployment of sparsified Ising machines~\cite{delacour2025self}.

\begin{figure*}[t]
    \centering
    \includegraphics[width=1\textwidth]{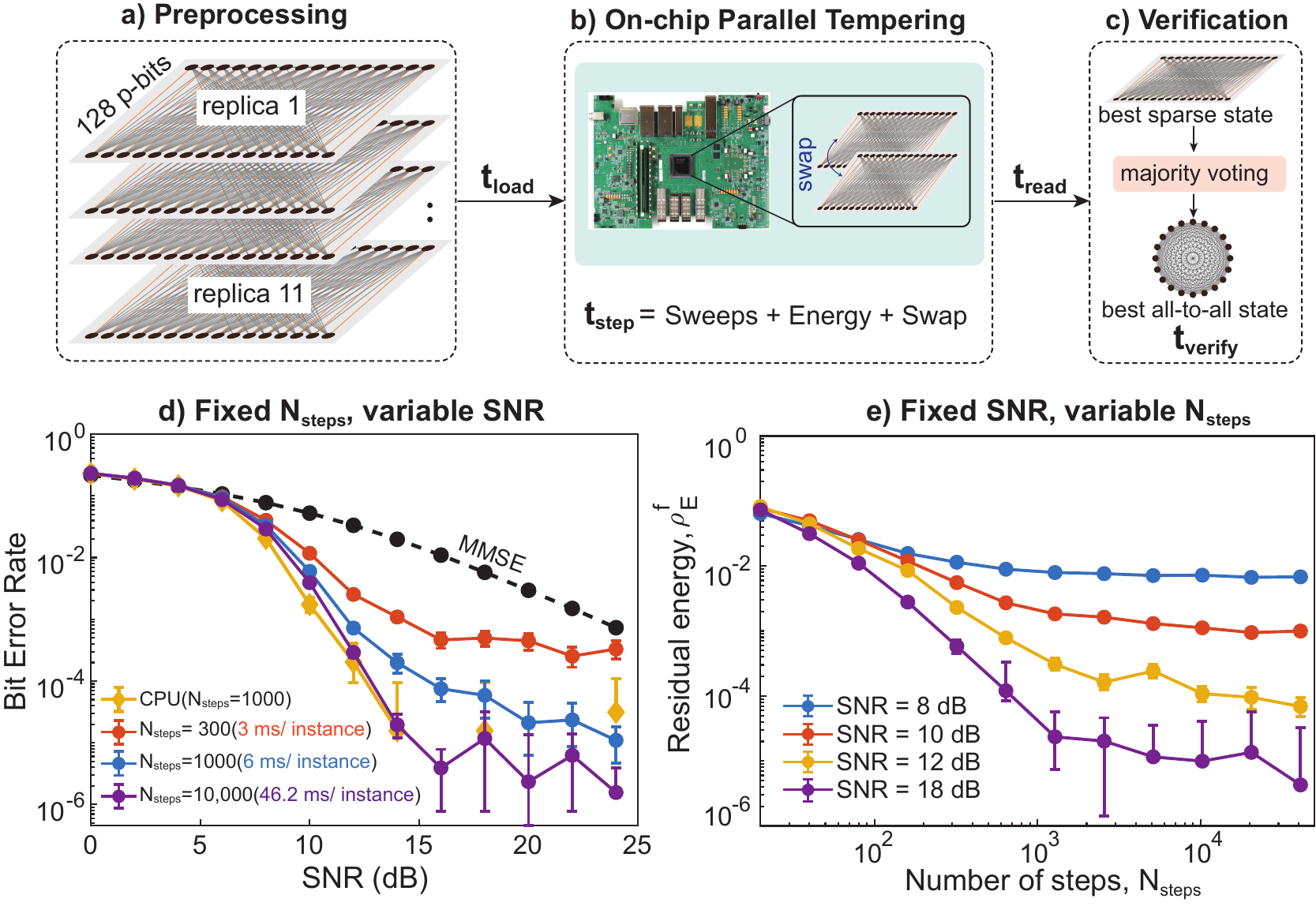}
    \caption{(a) Preprocessing: A 64-node all-to-all MIMO problem is sparsified to 128 p-bits (two copies per node) and replicated across 11 inverse temperatures ($\beta$ values), yielding a total of 1,408 p-bits for the 1D-PT architecture. The coupling weights ($J$) and biases ($h$) are loaded onto the FPGA in $t_{\mathrm{load}} = 0.76$~ms. (b) On-chip parallel tempering: Each step performs Monte Carlo sweeps, energy evaluation, and replica swaps, with a total step time of $t_{\text{step}}$. (c) Verification: The best sparse configuration is projected back to the original all-to-all graph via majority voting, read out in $t_{\text{read}} = 34.2\ \mu\text{s}$, and its energy is verified against the dense weight matrix in $t_{\text{verify}} = 213.5\ \mu\text{s}$. The complete solution time per instance is $t_{\text{instance}} \approx t_{\text{load}} + N_{\text{steps}} \times t_{\text{step}} + t_{\text{read}} + t_{\text{verify}}$, with timing values averaged over 130,000 instances. (d) Bit error rate (BER) versus SNR for $64 \times 64$ BPSK MIMO detection. The on-chip solver significantly outperforms the linear MMSE baseline. A CPU-based parallel tempering result ($N_{\text{steps}}=1000$) utilizing double precision is included for comparison, demonstrating that the fixed-point FPGA implementation maintains high algorithmic fidelity. The FPGA achieves complete end-to-end solution times of 3~ms ($N_{\text{steps}}=300$), 6~ms ($N_{\text{steps}}=1000$) and 46.2~ms ($N_{\text{steps}}=10,000$). (e) Residual energy $\rho_E^f = (E_{\text{meas}} - E_{\text{ground}})/N$ versus $N_{\text{steps}}$ for several SNR values. Error bars denote 95\% bootstrap confidence intervals.}
    \vspace{-5pt}
    \label{fig:fig2}
\end{figure*}

The rest of this paper develops the two methods in turn. Section~\ref{sec:onchip_pt} describes the fully on-chip 1D-PT solver for $64\times 64$ MIMO, fitting 11 temperature replicas of a 128-node sparsified system (1,408 p-bits) on an FPGA (FIG.~\ref{fig:fig1}c), with ASIC projections in 7~nm indicating 103~MHz at 285.8~mW. These projections should be viewed as a single-chip building block: problems exceeding one chip will still require decomposition or multi-chip coordination, but sparsification makes this building block compact and bounded-degree rather than all-to-all. On-chip experiments reveal a sharp sensitivity to the copy-constraint strength $P$, motivating Section~\ref{sec:2DPT}, where 2D-PT exchanges replicas across both temperature and constraint dimensions (FIG.~\ref{fig:fig1}d). Validation on the Sherrington-Kirkpatrick spin glass and on MIMO instances ($16\times 16$ in FPGA and $128\times 128$ in simulation) shows over $10\times$ faster convergence than 1D-PT with no manual parameter tuning, establishing a robust path toward scalable, tuning-free probabilistic computers.

\section{MIMO detection with p-bits}
\label{sec:mimo_scheme}

In a multiple-input multiple-output (MIMO) communication system, $N_t$ users each transmit a BPSK symbol $x_i \in \{-1,+1\}$, forming the vector $\mathbf{x} \in \{-1,+1\}^{N_t}$. The received signal is $\mathbf{y} = \mathbf{H}\mathbf{x} + \mathbf{n}$, where $\mathbf{H} \in \mathbb{R}^{N_r \times N_t}$ is the Rayleigh fading channel and $\mathbf{n}$ is additive white Gaussian noise. Maximum-likelihood detection seeks:
\begin{equation}
  \hat{\mathbf{x}}_{\text{ML}} = \arg\min_{\mathbf{x} \in \{-1,+1\}^{N_t}} \|\mathbf{y} - \mathbf{H}\mathbf{x}\|^2  
\end{equation}
Expanding and dropping constants, this maps to an Ising Hamiltonian $E(\mathbf{s}) = -\sum_i h_i s_i - \sum_{i<j} J_{ij} s_i s_j$ \cite{singh2022ising, singh2024uplink} with parameters
\begin{equation}
\mathbf{J} = -\mathbf{H}^T\mathbf{H}, \qquad \mathbf{h} = \mathbf{H}^T\mathbf{y}
\end{equation}
where any global prefactor is absorbed into the inverse temperature $\beta$. We solve this using p-bits updated according to:
\begin{equation}
I_i = \sum J_{ij} m_j + h_i, \qquad m_i = \mathrm{sgn}\bigl(\tanh(\beta I_i) - r_i\bigr)
\end{equation}
where $r_i \in [-1,1]$ is a uniform random number. This update rule samples from the Boltzmann distribution at inverse temperature $\beta$ \cite{camsari2017stochastic}. To accelerate convergence and escape local minima, we employ parallel tempering with multiple replicas spanning a range of $\beta$ values (Section~\ref{sec:onchip_pt}), and later extend to two-dimensional exchanges across both $\beta$ and constraint strength $P$ (Section~\ref{sec:2DPT}).

For reference, we compare against the minimum mean square error (MMSE) detector \cite{poor2000turbo, verdu1998multiuser}, $\hat{\mathbf{x}}_{\text{MMSE}} = (\mathbf{H}^T\mathbf{H} + \lambda \mathbf{I})^{-1}\mathbf{H}^T\mathbf{y}$, a standard linear baseline that balances interference suppression against noise amplification. Note that the unregularized limit ($\lambda$$\to$0) corresponds to continuously relaxing the Ising spins, illustrating how discrete optimization can outperform linear detection.

\begin{figure*}[t]
    \centering
    \includegraphics[width=1 \textwidth]{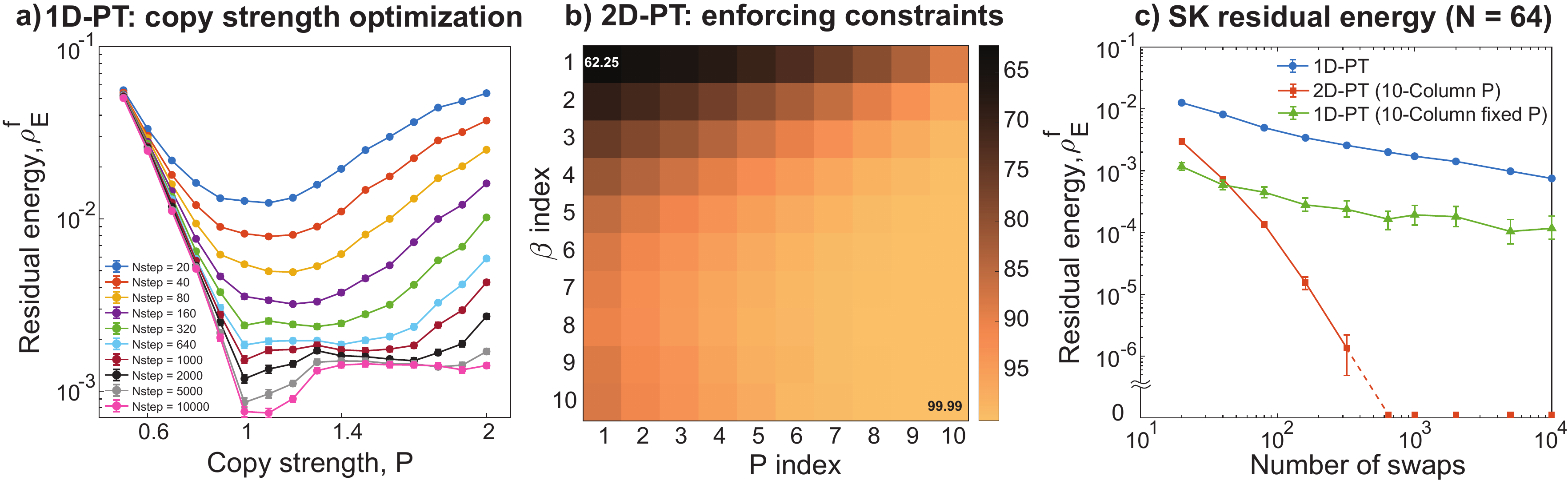}
   \caption{(a) Residual energy per spin $\rho_E^{f}$ for the Sherrington--Kirkpatrick (SK) spin glass ($N=64$, two-copy sparsification) under 1D parallel tempering as a function of copy strength $P$. (b) Copy-agreement percentage across the 2D replica grid ($\beta$ rows, $P$ columns) with $P \in [0.5, 2]$ reported in Table~\ref{tab:betaP_schedule}. The multi-column $P$ ladder enforces near-perfect agreement ($>$ 99\%) in the rightmost (high-$P$) column while allowing state exploration in low-$P$ columns. (c) Residual energy versus number of swaps. The 2D-PT approach (10-column, varying $P$) successfully reaches the ground state in fewer than $10^3$ swaps. In contrast, 1D-PT reduces energy at a significantly slower pace, requiring $10^4$ swaps to reach $\rho_E^f \approx 8\times 10^{-4}$, a value that 2D-PT achieves in just 40 swaps. For an iso-replica comparison, we evaluate a 10-column 1D-PT parallelization (selecting the best energy state across 10 independent 1D-PT columns at the optimal fixed $P$), which still fails to reach the ground states.  All data averaged over 100 instances $\times$ 100 trials, error bars denote 95\% bootstrap confidence intervals.}
    \label{fig:fig3}
    \vspace{-5pt}
\end{figure*}

\section{On-chip 1D parallel tempering (1D-PT)}
\label{sec:onchip_pt}

To map the dense MIMO problem onto hardware, we sparsify the all-to-all graph by introducing two auxiliary copies per node, connected by ferromagnetic couplings of strength $P$ (FIG.~\ref{fig:fig1}b). This yields 128 sparse p-bits from the original 64-node problem. We then implement Parallel Tempering (PT) \cite{swendsen1986replica,hukushima1996exchange} directly on an FPGA similar to designs in Ref. \cite{aadit2023accelerating, nikhar2024all}, with the distinction that all Monte Carlo sweeps, energy calculation, and Metropolis replica exchanges are entirely executed on-chip (FIG.~\ref{fig:fig2}b).

The sparse problem is expanded into an 11-temperature ladder, for a total of $128 \times 11 = 1{,}408$ p-bits instantiated concurrently (FIG.~\ref{fig:fig2}a). An optimized copy strength $P = 3.5$ is used to achieve the best bit-error rate for the computational budget. Weights are loaded once per channel with a time, $t_{\mathrm{load,J}} = 0.53$~ms, and biases are loaded for every symbol vector to be decoded per channel with a time, $t_{\mathrm{load,h}} = 0.23$~ms. So, the 100 channels and 100 symbols per channel will incur an effective loading time of $t_{\mathrm{load,h}} \approx 0.23$~ms, where the average $t_{\mathrm{load,J}}$ will be negligible. But considering the worst case scenario, all replicas are loaded once per instance with latency $t_{\mathrm{load}} =t_{\mathrm{load,J}}+t_{\mathrm{load,h}} = 0.76$~ms; after this one-time load, the dynamics proceed without further host communication. 

During the run, the lowest-energy replica at each swap is tracked, and the final readout corresponds to the best configuration encountered with $t_{\mathrm{read}} = 34.2$~$\mu$s. The sparse state is reduced to 64 logical p-bits via majority voting (ties resolved by coin flip) and verified against the original dense Ising Hamiltonian in $t_{\mathrm{verify}} = 213.5$~$\mu$s. Each parallel tempering step comprises Monte Carlo sweeps, energy computation, and a swap attempt:
\begin{equation}
 t_{\mathrm{step}} = t_{\mathrm{sweeps}} + t_{\mathrm{energy}} + t_{\mathrm{swap}} 
\end{equation}
The complete end-to-end time per instance is:
\begin{equation}
  t_{\mathrm{inst}} =  N_{\mathrm{steps}}\,t_{\mathrm{step}} + t_{\mathrm{overhead}}
\end{equation}

Here, $t_{\mathrm{overhead}} \approx t_{\mathrm{load,h}} + t_{\mathrm{read}} + t_{\mathrm{verify}} \approx 1$~ms, which is lower than experimentally measured overhead ($1.62$~ms which includes the instance generation). All timing values are averaged over 130,000 instances, and the reported solution times include all overheads: load, sweep, swap, read, and verify, with nothing hidden or neglected.

As shown in FIG.~\ref{fig:fig2}d, the on-chip solver achieves substantially lower bit error rates than the conventional MMSE detector. With $N_{\mathrm{steps}}=300$, 1000, and 10000 (100 sweeps per step), measured runtimes are 3~ms, 6~ms, and 46.2~ms per instance, respectively. Results are averaged over 100 independent channel realizations, each decoding 100 transmitted vectors at 13 SNR values (0 to 24dB in 2 dB steps). The residual energy (FIG.~\ref{fig:fig2}e) decreases systematically with $N_{\mathrm{steps}}$, saturating earlier at low SNR and following power-law scaling at high SNR.

The chip operates on the sparsified, undirected coupling matrix with a maximum degree $d_{\mathrm{max}} = 33$, and only the upper triangle of the symmetric weight matrix is transferred. With 10-bit weights and biases, each replica corresponds to $2080 \times 10 + 128 \times 10 = 22{,}080$ bits, so loading all 11 replicas requires $242{,}880$ bits ($\approx 30$~KB). At $t_{\mathrm{load}} = 0.76$~ms, the host-to-FPGA transfer rate is ${\approx}\,40$~MB/s, which is about $1.6\%$ of the $2.5$~GB/s available on PCIe. A majority of the speed loss is likely a result of the high-level MATLAB interface.

\begin{figure*}[t]
    \centering
    \includegraphics[width=0.85 \textwidth]{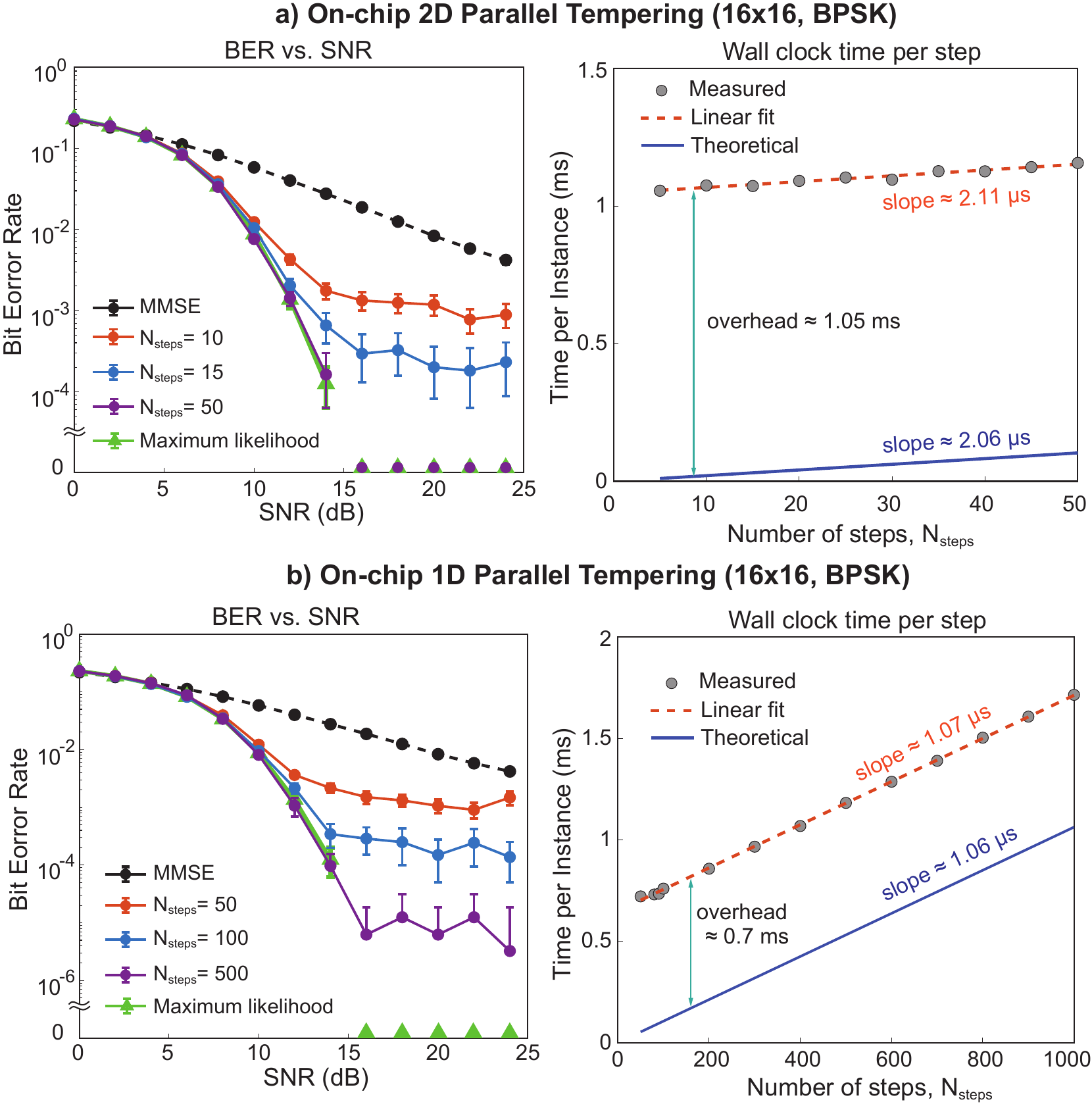}
    \caption{Performance and timing scaling for \textbf{on-chip} $16 \times 16$ BPSK MIMO detection, comparing Two-Dimensional (2D-PT) and One-Dimensional Parallel Tempering (1D-PT). (a) The 2D-PT architecture employs a grid of 54 total replicas (9 inverse temperature replicas $\times$ 6 copy strength replicas) and demonstrates rapid convergence, closely tracking the optimal Maximum Likelihood (ML) bound in just 50 steps. The measured wall clock time reveals an initial setup overhead of approximately 1.05 ms and a scaling slope of 2.11 $\mu\text{s}$ per step. (b) The 1D-PT architecture employs a standard ladder of 9 inverse temperature replicas with a fixed, optimized copy strength of $P=2.0$. Lacking the constraint exploration dimension, it suffers from an error floor (plateau) at high SNRs even after 500 steps, failing to reach the optimal ML performance. The corresponding timing plot shows a reduced initial overhead of about 0.7 ms and a slope of 1.07 $\mu\text{s}$ per step, reflecting the lower hardware utilization and algorithmic complexity compared to the full 2D grid.}
    \vspace{-10pt}
    \label{fig:fig4}
\end{figure*}

\section{2D Parallel Tempering}
\label{sec:2DPT}

The optimal coupling strength $P$ between copy nodes depends on the Monte Carlo budget and typically requires significant preprocessing. FIG.~\ref{fig:fig3}a shows the residual energy of 64-spin Sherrington-Kirkpatrick (SK) instances found by 1D Parallel Tempering (1D-PT) for various copy strengths and number of swaps (100 sweeps per swap). Regardless of the Monte Carlo budget, small copy strengths cannot find good solutions due to copy disagreement. Large copy strengths are ideal in theory as they enforce copy constraints. However, hard copy constraints hinder the search and require a large number of samples to find low-energy states, such as 10,000 swaps for reaching $\rho_E^f\approx 1.5\times 10^{-3}$ at $P=2$. In contrast, a similar energy residual is found for only 2,000 swaps at the optimal copy strength $P=1$, a 5$\times$ speedup compared to $P=2$. The optimal copy strength is a non-monotonic function of the swap budget, meaning that searching for the optimal is difficult and computationally expensive.

\begin{figure*}[t]
    \centering
    \includegraphics[width=0.85 \textwidth]{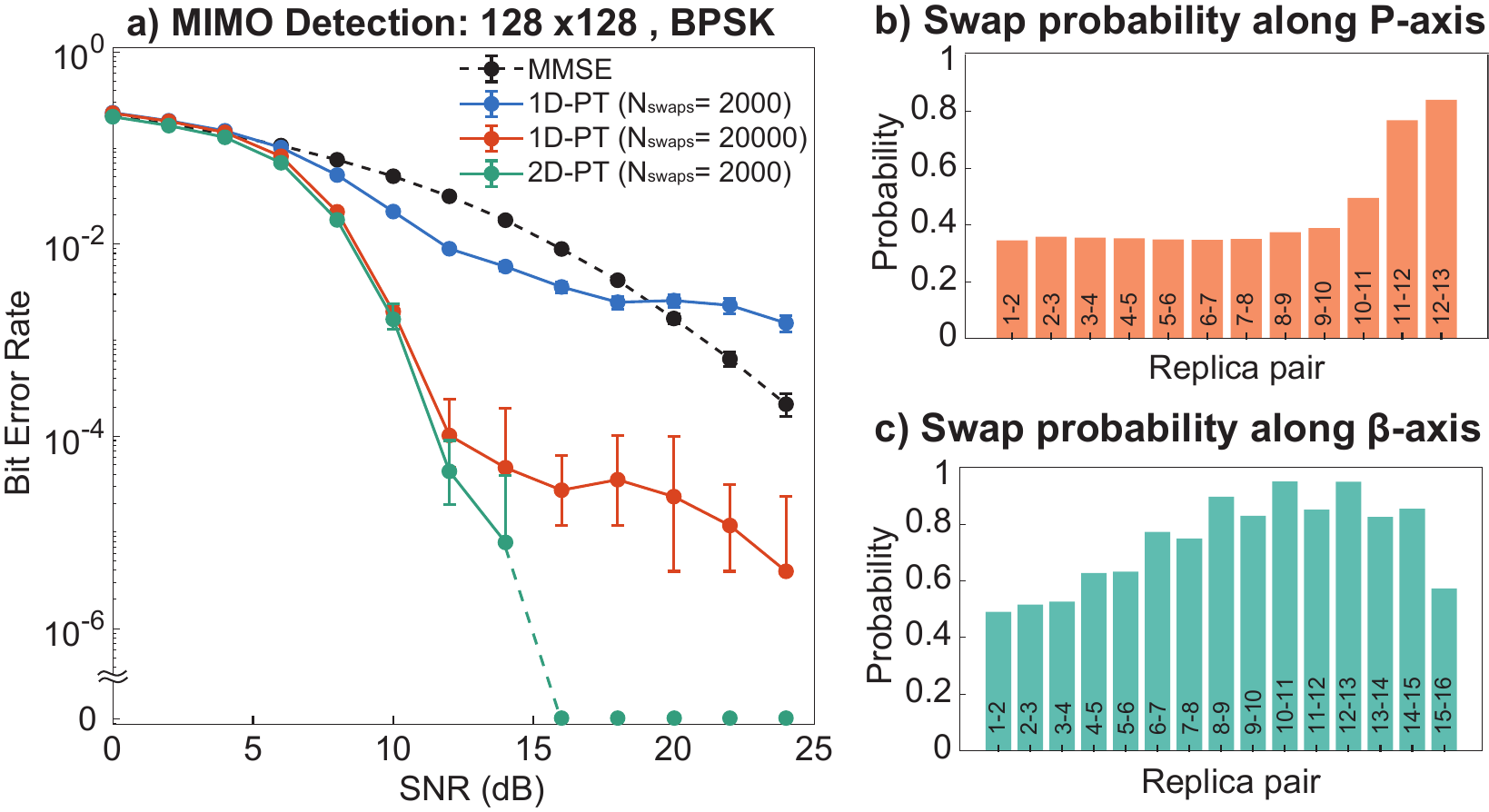}
    \caption{(a) Bit error rate (BER) versus signal-to-noise ratio (SNR) for $128\times128$ BPSK MIMO detection, averaged over 200 channels and 10 transmitted symbols per channel. All instances are sparsified with two copies per node (256 p-bits per replica). 1D-PT results are shown for two swap budgets ($N_{\mathrm{swaps}}=2,000$ and $20{,}000$) using a 16-replica $\beta$ ladder; increasing the budget by $10\times$ improves performance but 1D-PT still exhibits an error floor at high SNR. 2D-PT uses a $16\times13$ replica array with the same $\beta$ values and a distinct $P$ per column (values in Table~\ref{tab:betaP_schedule}), achieving substantially lower BER and reaching zero errors at high SNR. For both methods, the final solution is selected as the minimum-energy state after mapping back to the all-to-all problem; for 2D-PT, this selection is restricted to the coldest replicas (last row). (b,c) Mean swap probabilities along the $P$-axis and $\beta$-axis, respectively, averaged over all rows and columns.}
    \label{fig:fig5}
\end{figure*}

Instead of co-optimizing penalty strength and Monte Carlo budget, we use Two-Dimensional Parallel Tempering (2D-PT) \cite{delacour2025two}, which automatically enforces hard constraints while benefiting from fast-mixing replicas having small to medium penalty strengths. As illustrated in FIG.~\ref{fig:fig1}d, 2D-PT explores various copy strengths in a new dimension dedicated to constraint exploration, orthogonal to the inverse temperature axis ($\beta$), resulting in an array of replicas. Each column of the 2D array has a fixed penalty strength $P$, interpolating from soft constraints in the first column to hard constraints in the last one. In addition to vertical configuration exchanges along the inverse temperature ($\beta$-swaps), new horizontal exchanges in the penalty direction ($P$-swaps) transfer feasible states found in the left-most columns (fast mixing) to the right-most columns that enforce constraints (slow mixing).  FIG.~\ref{fig:fig3}b shows the $10\times10$ array used for the next SK experiments, where the last-column bottom replica enforces a mean copy-agreement of 99.99\% and holds the lowest-energy solution during the search.

We first apply 2D-PT on 64-spin (128-spin sparsified) SK instances, which, like MIMO, are all-to-all graphs with Gaussian couplings but have no noise or external fields. The 2D-PT $\beta$ and $P$ schedules are found with the adaptive algorithm described in the Method Section and reported in Table~\ref{tab:betaP_schedule}. 1D-PT employs the same ten $\beta$ values as 2D-PT. To ensure a fair comparison, each 1D-PT data point from FIG.~\ref{fig:fig3}c uses the copy strength $P$ optimized for that specific swap budget (provided by FIG.~\ref{fig:fig3}a), resulting in the best possible 1D-PT configuration for the given $\beta$ schedule. 
Due to limited FPGA capacity, the results in FIG.~\ref{fig:fig3}c reflect 1D-PT executed on the FPGA and 2D-PT simulated in MATLAB. This cross-platform comparison is valid because the FPGA-based 1D-PT results closely match a full-precision baseline (see Supplementary Material). 
While 1D-PT requires approximately $10^4$ swaps to achieve a mean residual energy of $\rho_E^f \approx 8 \times 10^{-4}$, 2D-PT reaches this same energy level in roughly 40 swaps, yielding a $250\times$ speedup. FIG.~\ref{fig:fig3}c also compares 2D-PT against a 10-column 1D-PT parallelization using the same total number of replicas (where the minimum energy is selected across all 10 columns). This iso-replica comparison demonstrates that 2D-PT does not merely trade space for runtime; parallelizing 1D-PT across 10 independent columns fails to match 2D-PT's algorithmic advantage. Most notably, 2D-PT finds the constrained ground states ($\rho_E^f=0$) in fewer than $10^3$ swaps, a regime that remains inaccessible to the 10-column 1D-PT approach even with a budget of $10^4$ swaps.

\subsection{On-chip Two-Dimensional Parallel Tempering (2D-PT)}

Motivated by these algorithmic gains, we evaluate the practical advantages of exploring the constraint dimension in hardware by implementing and comparing both 1D-PT and 2D-PT architectures for a $16 \times 16$ BPSK MIMO detection problem entirely on-chip. In FIG.~\ref{fig:fig4}, the 2D-PT solver utilizes an expanded grid of 54 parallel replicas, arranged orthogonally across 9 inverse temperatures ($\beta$) and 6 copy strengths ($P$). This two-dimensional state space allows the system to rapidly escape local minima by migrating feasible configurations through varying constraint rigidities. Consequently, the 2D-PT architecture demonstrates remarkable algorithmic efficiency, closely tracking the optimal Maximum Likelihood (ML) bound across the entire SNR range in just $N_{\text{steps}} = 50$. The traditional 1D-PT architecture, restricted to a 9-replica temperature ladder with a fixed, optimized penalty of $P = 2.0$ suffers from an error floor at high SNRs, failing to reach the optimal ML performance even after $N_{\text{steps}} = 500$.

While the algorithmic convergence of 2D-PT is vastly superior, the expanded 54-replica grid introduces expected tradeoffs in hardware area. FIG.~\ref{fig:fig4} also outlines the measured wall-clock execution time per instance. Because the 2D-PT implementation must initialize and manage significantly more p-bits and inter-replica swap controllers, it incurs a slightly higher initial setup overhead of approximately $1.05\,\text{ms}$, compared to $0.7\,\text{ms}$ for the leaner 1D-PT system. Furthermore, the per-step execution latency is higher for the 2D grid, scaling at a measured slope of $2.11\,\mu\text{s}$ per step versus $1.07\,\mu\text{s}$ per step for the 1D ladder, due to the secondary swap direction.

\subsection{Scaling 2D-PT for $128\times 128$ MIMO}
To assess how 2D-PT scales beyond the on-chip $16\times 16$ demonstration, we apply it to $128\times 128$ BPSK MIMO instances in MATLAB, using a $16\times 13$ replica array (16 $\beta$ values, 13 $P$ values) with 5 sweeps per swap; schedules are given in Table~\ref{tab:betaP_schedule}. For 2,000 swaps, 2D-PT significantly outperforms MMSE across the entire SNR range, with fully decoded signals (BER=0) at large SNR values ($>$15 dB). In contrast, 1D-PT barely improves MMSE for the same Monte Carlo budget and has a larger BER at high SNR as shown in FIG.~\ref{fig:fig5}a. When increasing the number of swaps to 20,000, 1D-PT recovers 2D-PT's performance in noisy regimes (low SNR), but cannot perfectly decode all instances at high SNR (error floor of around $10^{-5}$). Beyond accelerating MIMO decoding by more than $10\times$ compared to 1D-PT, 2D-PT has the additional advantage of not requiring tuning of the penalty strength for every Monte Carlo budget (as in FIG.~\ref{fig:fig3}a).

2D-PT's only requirement for high performance is to enable some state swaps between adjacent replica pairs, acting as global moves to escape local minima. The exchange probability between replicas depends on the $\beta$ and $P$ schedules, obtained by averaging individual schedules from 10 random instances and reported in Table~\ref{tab:betaP_schedule}.  The high swap probabilities in FIG.~\ref{fig:fig5}b confirm efficient mixing via frequent state exchanges along the $P$-axis, enabling feasible states to migrate into hard-constraint replicas. The $\beta$-axis swap probability remains high across all temperature pairs (FIG.~\ref{fig:fig5}c), transferring low-energy solutions to the final $\beta$ row.

\begin{figure*}[t]
\vspace{-10pt}
    \centering
    \includegraphics[width=1\textwidth]{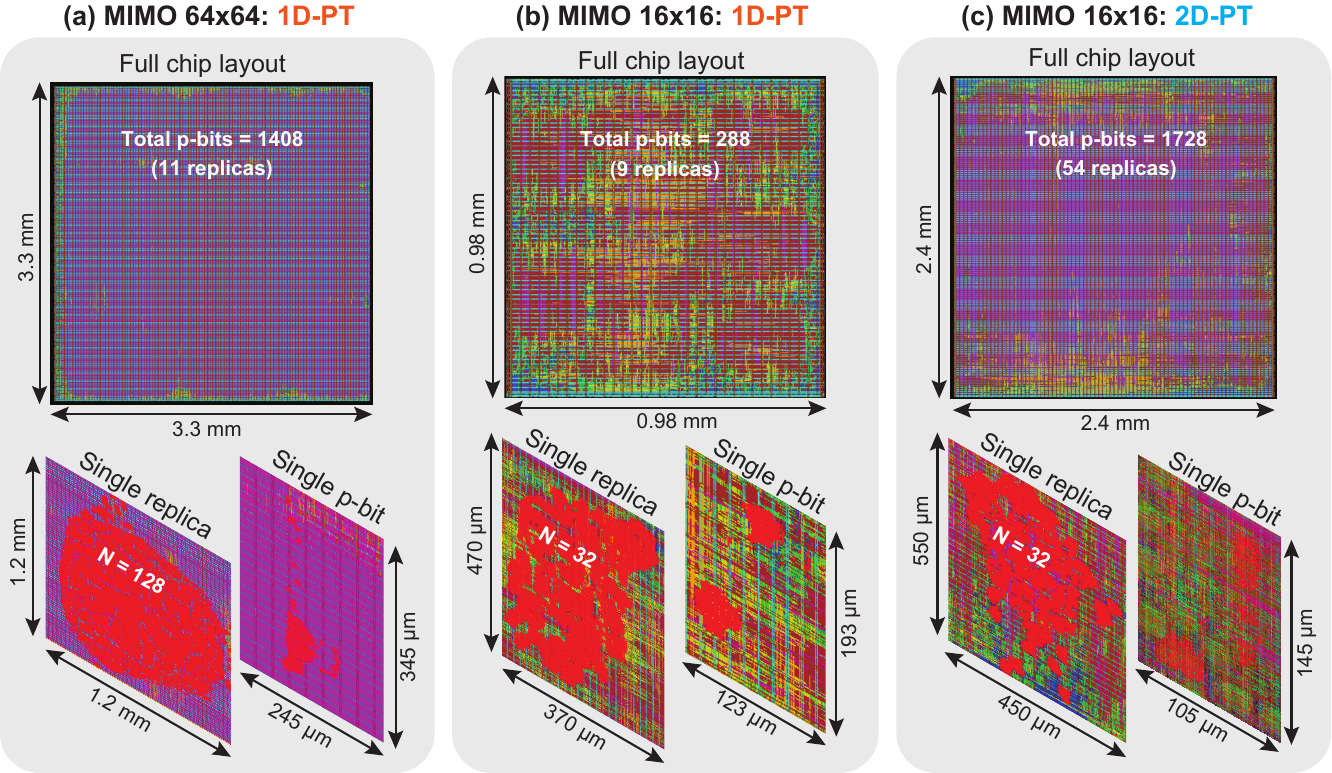}
    \caption{Graphic Database System II (GDSII) layouts for Parallel Tempering (PT) architectures solving Multiple-Input Multiple-Output (MIMO) problems. All designs (a)-(c) follow a three-level hierarchy showing the full chip layout, a single replica macro, and an individual p-bit macro, with all layouts synthesized for a target frequency of 100~MHz. (a) 1D Parallel Tempering mapped to a 64 $\times$ 64 MIMO problem ($N = 128$ p-bits per replica, 11 replicas). The top-level floorplan integrates 11 parallel replicas (1,408 total p-bits), occupying a total die area of 10.9~mm$^2$ (3.3 $\times$ 3.3~mm). The single replica macro occupies 1.2 $\times$ 1.2~mm, and an individual p-bit macro is 245 $\times$ 345~$\mu$m. The total estimated power for this design is 285.8~mW with 4M gate instances. (b) 1D Parallel Tempering mapped to a 16 $\times$ 16 MIMO problem ($N = 32$ p-bits per replica, 9 replicas). The top-level floorplan integrates 9 parallel replicas (288 total p-bits), occupying a total die area of 0.96~mm$^2$ (0.98 $\times$ 0.98~mm). The single replica macro occupies 470 $\times$ 370~$\mu$m, and an individual p-bit macro is 123 $\times$ 193~$\mu$m. The total estimated power for this design is 19.6~mW with 371K gate instances. (c) 2D Parallel Tempering mapped to a 16 $\times$ 16 MIMO problem ($N = 32$ p-bits per replica, 54 replicas). The top-level floorplan integrates 54 parallel replicas (1,728 total p-bits), occupying a total die area of 5.76~mm$^2$ (2.4 $\times$ 2.4~mm). The single replica macro occupies 550 $\times$ 450~$\mu$m, and an individual p-bit macro is 145 $\times$ 105~$\mu$m. The total estimated power for this design is 124~mW with 2.3M gate instances.}
    \label{fig:figASIC}
    \vspace{-5pt}
\end{figure*}

\begin{table}[b]
\vspace{-10pt}
\centering
\caption{Key metrics of synthesized p-computer designs: maximum frequency, chip area, and total expected power usage.}
\label{tab:asics}
\renewcommand{\arraystretch}{1.2}
\setlength{\tabcolsep}{4pt}   

\resizebox{\columnwidth}{!}{%
\begin{tabular}{lccccc}
\toprule
\textbf{Design Size} & \textbf{\shortstack{No. of\\P-bits}} & \textbf{\shortstack{Expected\\Freq. (MHz)}} & \textbf{\shortstack{Total Area\\(mm$^2$)}} & \textbf{\shortstack{Total Power\\(mW)}}  & 
\textbf{\shortstack{Total Gate\\ Instances}} \\
\midrule
$64 \times 64$, 1D-PT & 1408 & 103  & 10.9  & 285.8 & 4M \\ 
$16 \times 16$, 1D-PT & 288 & 157  & 0.96  & 19.6 & 371K \\ 
$16 \times 16$, 2D-PT & 1728 & 111  & 5.76  & 124 & 2.3M \\ 
\bottomrule
\end{tabular}%
}
\vspace{-10pt}
\end{table}

\section{ASIC Physical Design \& Implementation}
\label{sec:asic}

To project the performance of our solver to a dedicated hardware platform, we targeted an Application-Specific Integrated Circuit (ASIC) implementation of the same $64 \times 64$ 1D, $16 \times 16$ 1D, and $16 \times 16$ 2D BPSK MIMO instance evaluated on the FPGA. We explored the configuration with two copies per node using the ASAP7 predictive Process Design Kit (PDK) having 9 metal layers ~\cite{clark2016asap7}. To manage the complexity of this large-scale system, we employed a toolchain called \emph{mflowgen} \cite{carsello2022mflowgen}, which builds upon the Cadence genus/innovus flow, to manage the synthesis of the full design topology. Note that metrics in ASAP7 through the mflowgen/innovus toolchain are not representative of true 7nm technology, and are area-wise 28nm technology, but electrically 7nm technology. This top-level design was fully synthesized to sign off, including a realistic clock-tree, placement, and routing, to ensure the rigorous characterization of area, timing, and power metrics. The designs are targeting 70\% density. The designs are signoff-quality at the typical corner and close timing above $100$~MHz. The operating frequencies reported in Table~\ref{tab:asics} are extracted from the worst-negative slack.

The $64 \times 64$ 1D-PT BPSK MIMO system, sparsified with two copies per node and instantiating 11 parallel replicas, achieves an operating frequency of 103\,MHz with a total power consumption of 285.8\,mW (110\,mW internal, 174.6\,mW switching, 0.51\,mW leakage) and an effective logic area of 7.6\,mm$^2$. Where effective logic area refers to the total area occupied by the computation logic, which differs from the total die area, that includes power, pins, and other overhead present in the layout.

The $16 \times 16$ 1D-PT BPSK MIMO system, sparsified with two copies per node and instantiating 9 parallel replicas, achieves an operating frequency of 157\,MHz with a total power consumption of 19.6\,mW (9.9\,mW internal, 9.7\,mW switching, 0.04\,mW leakage) and an effective logic area of 0.65\,mm$^2$. The same system using 2D-PT and instantiating 54 parallel replicas achieves an operating frequency of 111\,MHz with a total power consumption of 124\,mW (59.6\,mW internal, 64\,mW switching, 0.25\,mW leakage) and an effective logic area of 3.9\,mm$^2$. All area metrics are presented directly using the scale provided by the PDK to the toolchain. 

Power metrics are defined with the total static and dynamic at 50\% gate activity/switching factor and 10\% input activity. The 50\% activity factor was chosen based on the worst-case number of p-bits switching in our 3-coloring graph.
The corresponding fully integrated physical layouts (GDS mask) are illustrated in FIG~\ref{fig:figASIC}. Results for the 1D-PT two-copy and 2D-PT two-copy configurations are summarized in Table~\ref{tab:asics}. It is important to note that \texttt{Single replica} and \texttt{Single p-bit} sizes do not expressly represent hierarchical size scaling. Due to overlapping area between multiple p-bits and multiple replicas, sublinear scaling between hierarchy tiers is observed. At the system level, PT is not naturally pipelined within a single Markov chain, so higher aggregate throughput must come from coarse-grained parallelism across independent PT engines, as quantified in Tables~S2 and~S3.

\section{Conclusion}
\label{sec:conclusion}

We demonstrated a fully on-chip probabilistic computer for dense combinatorial optimization, benchmarked on MIMO detection. By sparsifying the problem graph and implementing parallel tempering on an FPGA, our $64 \times 64$ 1D-PT architecture (1,408 p-bits) achieves lower bit error rates than linear detectors with 3~ms end-to-end solution times. ASIC projections for this design indicate 103~MHz operation at 285.8~mW.

To eliminate the manual copy-constraint tuning required by 1D-PT, we utilize Two-Dimensional Parallel Tempering (2D-PT) to dynamically exchange replicas across both temperature and constraint dimensions. 2D-PT converges roughly $250\times$ faster on SK spin glasses and achieves zero bit errors at high SNRs for massive $128 \times 128$ MIMO instances, effectively eliminating 1D-PT's error floor. We validated this algorithmic advantage in hardware with a fully on-chip $16 \times 16$ 2D-PT solver. By utilizing an expanded grid of 54 replicas (1,728 p-bits), it achieved highly accurate decoding in just 50 steps, compared to 500 steps for a standard 1D-PT ladder. With ASIC projections showing 111~MHz and 124~mW operation, these results establish a robust, tuning-free algorithmic framework and a scalable on-chip architecture capable of meeting the throughput demands of next-generation wireless systems.

\begin{table*}[t]
\centering
\caption{2D-PT parameters for the SK ($N=64$) and MIMO ($16 \times16$ and $128 \times 128$, BPSK) experiments, where both problems are sparsified with 2 copies per original node. $\alpha_{\beta}$ and $\alpha_{P}$ are the parameters used in the adaptive algorithm to obtain the $\beta$ and $P$ values iteratively \cite{delacour2025two}.}
\label{tab:betaP_schedule}

\small
\setlength{\tabcolsep}{2pt} 
\renewcommand{\arraystretch}{1.25} 

\resizebox{\textwidth}{!}{%
    \begin{tabular}{c c c *{16}{c}} 
    \toprule
    \textbf{\shortstack{Problem}} & \textbf{\shortstack{Param.}} & \textbf{\shortstack{Row/Column}}  &
    \textbf{1}&\textbf{2}&\textbf{3}&\textbf{4}&\textbf{5}&\textbf{6}&\textbf{7}&\textbf{8}&
    \textbf{9}&\textbf{10}&\textbf{11}&\textbf{12}&\textbf{13}&\textbf{14}&\textbf{15}&\textbf{16}\\
    \midrule

    \multirow{2}{*}{\shortstack{SK\\($N=64$)}} 
     & $\alpha_\beta=2.5$& $\beta$-value & 0.500 & 0.801 & 1.13 & 1.52 & 2.05 & 2.92 & 4.62 & 8.60 & 24.6 & 138 & -- & -- & -- & -- & -- & --\\
     & $\alpha_P=0.4$ & $P$-value      & 0.500 & 0.572 & 0.649 & 0.740 & 0.846 & 0.970 & 1.12 & 1.33 & 1.59 & 2.05 & -- & -- & -- & -- & -- & -- \\
    \midrule

    \multirow{2}{*}{\shortstack{MIMO\\($16\times 16$)}} 
     & $\alpha_\beta=1.25$& $\beta$-value & 0.500 & 0.760 & 1.10 & 1.55 & 2.22 & 3.31 & 5.34 & 10.3 & 27.9 & -- & -- & -- & -- & -- & -- & --\\
     & $\alpha_P=0.75$ & $P$-value      & 0.100 & 0.357 & 0.658 & 1.02 & 1.50 & 2.26 & -- & -- & -- & -- & -- & -- & -- & -- & -- & -- \\
    \midrule

    \multirow{2}{*}{\shortstack{MIMO\\($128\times 128$)}} 
     & $\alpha_\beta=1.25$ & $\beta$-value & 0.500 & 0.564 & 0.647 & 0.743 & 0.859 & 1.00 & 1.18 & 1.40 & 1.71 & 2.16 & 2.88 & 4.24 & 7.46 & 18.0 & 72.0 & 332 \\
     & $\alpha_P=0.75$ & $P$-value      & 0.800 & 0.970 & 1.15 & 1.33 & 1.54 & 1.76 & 2.02 & 2.32 & 2.69 & 3.21 & 3.96 & 4.85 & 9.44 & -- & -- & -- \\
    \bottomrule
    \end{tabular}%
}
\end{table*}

\appendix

\section{Methods}
\label{sec:methods}
\subsection{FPGA design}
To evaluate the on-chip 1D and 2D parallel tempering solvers, we implemented the designs on an Alveo U250 FPGA accelerator card, which interfaces with a host CPU via PCIe for initialization, control, and readout. The coupling weights $J_{ij}$ and biases $h_i$ (pre-multiplied by the inverse temperature $\beta$) are represented using a 10-bit fixed-point precision format (Q6.3). For the 1D-PT architecture targeting a $64 \times 64$ BPSK MIMO configuration, we instantiated 11 parallel replicas. Each replica contains 128 p-bits representing a sparsified $N=64$ graph with two copies per node, resulting in a total of 1,408 on-chip p-bits. Furthermore, our 2D-PT implementation for a $16 \times 16$ BPSK MIMO problem required an expanded grid of 54 replicas, totaling 1,728 on-chip p-bits. 

The on-chip probabilistic computer is built using a modular SystemVerilog architecture that natively executes the Two-Dimensional Parallel Tempering (2D-PT) algorithm. 2D-PT organizes these instantiated replicas into a two-dimensional grid where rows correspond to varying inverse temperatures ($\beta$), and columns correspond to varying copy-constraint strengths ($P$). A central Finite State Machine (FSM) orchestrates the execution flow, sequentially interleaving parallel Monte Carlo sweeps with Metropolis replica swap attempts in both the $\beta$ and $P$ directions.

To achieve this massive parallelism, the hardware architecture strictly minimizes its logic footprint. The fundamental computing unit, the p-bit, utilizes a 32-bit Galois linear feedback shift register (LFSR) for independent pseudo-random number generation and a lookup table to evaluate the $\tanh$ activation function. Crucially, the design minimizes the need for hardware multipliers (DSP blocks) wherever possible. Because p-bit states are binary, multiplications are replaced by conditional sign-selection operations and summed via parallel binary adder trees. The graph colors non-adjacent p-bits into independent sets that update simultaneously within a single clock cycle. Furthermore, all replica exchange probabilities are calculated entirely on-chip using a low-latency, base-2 exponential approximation of the Metropolis criterion. 

Comprehensive details regarding the adaptive $\beta$ and $P$ scheduling algorithm, block diagrams, FSM timing characteristics, and complete FPGA resource utilization metrics are provided in the \textbf{Supplementary Information}.

\subsection{2D-PT parameters}
The $\beta$ and $P$ schedules for 2D-PT are determined iteratively using the adaptive algorithm proposed in Ref. \cite{delacour2025two}. Beginning with the initial $\beta$ and $P$ values at the top-left replica, the parameter grid is expanded according to:
\begin{align}
    \beta(i+1,j) & =\beta(i,j)+\alpha_\beta/ \sigma_E \\ \nonumber
    P(i,j+1)&=P(i,j)+\alpha_P/(\beta(i,j) \sigma_g)
\end{align}
Here, $i$ and $j$ denote the indices along the inverse temperature ($\beta$) and constraint ($P$) dimensions, respectively, while $\alpha_\beta$ and $\alpha_P$ control the step size between consecutive values. The terms $\sigma_E$ and $\sigma_g$ represent the standard deviations of the energy and the constraint function ($g=0$ indicating satisfied constraints) for replica $(i,j)$, estimated using 20 independent chains of $10^3$ Monte Carlo sweeps. The adaptive algorithm stops row expansion when $\sigma_E$ falls below a defined threshold (one-tenth of the mean weight amplitude) and halts column expansion once all samples satisfy the constraints. The final $\beta$ and $P$ vectors are derived from the median values of the arrays along the column and row directions, respectively. The step parameters $\alpha_\beta$ and $\alpha_P$ were initially tuned on a single instance to ensure a minimum of 10 values in each direction. For the experiments shown in FIG.~\ref{fig:fig4} and FIG.~\ref{fig:fig5}, the final schedules are averaged over 10 instances of SK and MIMO problems (Table~\ref{tab:betaP_schedule}). In all cases, the single-column 1D-PT employs the same $\beta$ schedule as 2D-PT.

\section*{Acknowledgment}
MMHS, KC-C, CD, and KYC acknowledge support from the Office of Naval Research (ONR), Multidisciplinary University Research Initiative (MURI) grant N000142312708. MMHS and KYC acknowledge support from the Semiconductor Research Corporation (SRC) grant. SS and TS acknowledge support from CMU CIT dean's fellowship.

\section*{Data availability}
The data that support the findings of this article are not publicly available. The data are available from the authors upon reasonable request.

\section*{Code availability}
The computer code used in this study is available from the corresponding author upon reasonable request.

\section*{Author contributions}
MMHS and KYC conceived the study. KYC supervised the study. MMHS and KC-C performed on-chip (FPGA) 1D-PT and 2D-PT experiments, respectively. CD conducted 2D-PT experiments in simulation. KC-C developed RTL code to implement on-chip PT. SS and TS implemented the ASIC physical design. All authors discussed and analyzed the experiments and participated in writing the manuscript. 

\section*{Competing interests}
The authors declare no competing interests.

\bibliographystyle{unsrtnat}

\balance

\pagebreak
\clearpage
\newpage


\setcounter{secnumdepth}{3}

\newcommand{\beginsupplement}{%
        \ifdefined\counterwithout
            \counterwithout{equation}{section}
        \fi
        \setcounter{table}{0}
        \renewcommand{\thetable}{S\arabic{table}}%
        \setcounter{figure}{0}
        \renewcommand{\thefigure}{S\arabic{figure}}%
        \setcounter{equation}{0}
        \renewcommand{\theequation}{S.\arabic{equation}}%
        \renewcommand{\thealgocf}{S\arabic{algocf}}
        \setcounter{algocf}{0}%
     }

\onecolumngrid

\newcounter{rsection}
\newcommand{\rsection}[1]{%
  \refstepcounter{rsection}%
  \vspace{14pt}%
  \begin{center}%
    \normalfont\large\bfseries\Roman{rsection}.\enspace #1%
  \end{center}%
  \vspace{3pt}%
}

\renewcommand{\thesubsection}{\arabic{subsection}}
\renewcommand{\subsection}[1]{%
  \refstepcounter{subsection}
  \setcounter{subsubsection}{0}%
  \vspace{8pt}%
  \noindent{\normalfont\large\thesubsection.\enspace \textit{#1}}%
  \par\vspace{3pt}%
}

\renewcommand{\thesubsubsection}{\thesubsection.\arabic{subsubsection}}
\renewcommand{\subsubsection}[1]{%
  \refstepcounter{subsubsection}
  \vspace{6pt}%
  \noindent{\normalfont\large\thesubsubsection\enspace #1}%
  \par\vspace{3pt}%
}

\newcounter{algcounter}
\newenvironment{algbox}[1]{%
  \refstepcounter{algcounter}%
  \begin{mdframed}[
    linecolor=black,
    linewidth=0.8pt,
    topline=true, bottomline=true,
    leftline=false, rightline=false,
    innerleftmargin=4pt, innerrightmargin=4pt,
    innertopmargin=6pt, innerbottommargin=6pt,
    backgroundcolor=white,
    nobreak=true
  ]
  \noindent\textbf{Algorithm~S\thealgcounter:} #1\\[3pt]
  \hrule\vspace{5pt}
}{\end{mdframed}}

\newcommand{\KW}[1]{\textbf{#1}}
\newcommand{\CMT}[1]{\textit{/* #1 */}}
\newcommand{\IND}{\hspace{1.0em}}

\setlist[itemize]{leftmargin=1.5em,itemsep=2pt,topsep=4pt}
\setlistdepth{9}
\renewlist{enumerate}{enumerate}{9}
\setlist[enumerate,1]{label=\arabic*.}
\setlist[enumerate,2]{label=\alph*.}
\setlist[enumerate,3]{label=\roman*.}
\setlist[enumerate,4]{label=\Alph*.}
\setlist[enumerate,5]{label=\Roman*.}
\setlist[enumerate,6]{label=\arabic*.}

\setlength{\parskip}{3pt}
\setlength{\parindent}{1.5em}

\renewcommand{\thefigure}{S\arabic{figure}}
\renewcommand{\thetable}{S\arabic{table}}
\renewcommand{\theequation}{S\arabic{equation}}

\begin{center}
  {\normalfont\Large\bfseries SUPPLEMENTARY INFORMATION}\\[8pt]
  {\normalfont\normalsize\bfseries Probabilistic Computers for MIMO Detection:
  From Sparsification to 2D Parallel Tempering}\\[6pt]
  {\small
  M Mahmudul Hasan Sajeeb,$^{1,*}$
  Kevin Callahan-Coray,$^{1,*}$
  Corentin Delacour,$^{1}$\\
  Sanjay Seshan,$^{2}$
  Tathagata Srimani,$^{2}$ and
  Kerem Y. Camsari$^{1}$}\\[5pt]
  {\small\itshape
   $^{1}$Department of Electrical and Computer Engineering,
   University of California, Santa Barbara, CA 93106, USA}\\
  {\small\itshape
   $^{2}$Department of Electrical and Computer Engineering,
   Carnegie Mellon University, Pittsburgh, PA 15213, USA}\\
  {\small\itshape
   $^{*}$Equally contributing authors}\\
\end{center}

\crefname{subsection}{Sec.}{Secs.}
\Crefname{subsection}{Section}{Sections}
\crefname{subsubsection}{Sec.}{Secs.}
\Crefname{subsubsection}{Section}{Sections}
\crefname{equation}{Eq.}{Eqs.}
\Crefname{equation}{Equation}{Equations}
\crefname{table}{Table}{Tables}
\Crefname{table}{Table}{Tables}
\crefname{figure}{Fig.}{Figs.}
\Crefname{figure}{Figure}{Figures}

\makeatletter
\renewcommand{\p@subsection}{}
\renewcommand{\p@subsubsection}{}
\makeatother

\crefalias{subsection}{subsection}
\crefalias{subsubsection}{subsubsection}

\newcommand{\ms}{\,\text{ms}}
\newcommand{\mus}{\,\mu\text{s}}
\newcommand{\ns}{\,\text{ns}}
\newcommand{\MHz}{\,\text{MHz}}
\newcommand{\kbps}{\,\text{kbps}}
\newcommand{\uJpb}{\,\mu\text{J/bit}}
\newcommand{\mJpb}{\,\text{mJ/bit}}

\beginsupplement
\renewcommand{\theHfigure}{S\arabic{figure}}
\renewcommand{\theHtable}{S\arabic{table}}
\renewcommand{\theHequation}{S.\arabic{equation}}
\renewcommand{\theHalgocf}{S\arabic{algocf}}
\renewcommand{\theHsection}{S\arabic{section}}

\setcounter{section}{0}
\setcounter{subsection}{0}

\vspace{-15pt}

\rsection{FPGA Architecture for On-chip ( 1D and 2D ) Parallel Tempering}

This section provides a detailed hardware description of the on-chip Parallel Tempering (PT) architecture, covering both the one-dimensional (1D-PT) and two-dimensional (2D-PT) implementations. The design is implemented in SystemVerilog and targets FPGA deployment, with the architecture being directly portable to ASIC flows. The 1D-PT and 2D-PT designs share the same modular hierarchy; the only structural difference is that 1D-PT arranges replicas in a single column (column count $C = 1$) with the finite state machine (FSM) executing only $\beta$-swap loops, while 2D-PT arranges replicas in an $R \times C$ grid and the FSM alternates between $\beta$-swap and $P$-swap (constraint-swap) loops. All other modules, the p-bit, the synapse (MAC and local energy), the tanh lookup table, the LFSR, the replica, the energy accumulator, the infeasibility counter, the swap controllers, and the readout logic, are identical between the two configurations.

The remainder of this section describes each module in the design hierarchy from the bottom up, followed by a discussion of the FSM that orchestrates the parallel tempering algorithm, and concludes with the top-level integration and communication infrastructure.

\subsection{Setup}

\noindent{\textbf{Communication:}} All data transfer between the FPGA and host device is conducted over a PCIe connection (Gen 3, 8 lanes). Data movement inside the FPGA is handled via the AXI protocol and is controlled by the MathWorks' PCIe Express AXI Manager IP. The communication side of the FPGA operates on a separate clock domain fixed at 125\,MHz by the PCIe DMA/Bridge module. Because this frequency is not guaranteed to be achievable by the PT datapath, SmartConnect modules handle the clock domain crossing between the communication and PT clock domains.

\begin{figure}[H]
\centering
\includegraphics[width=\textwidth]{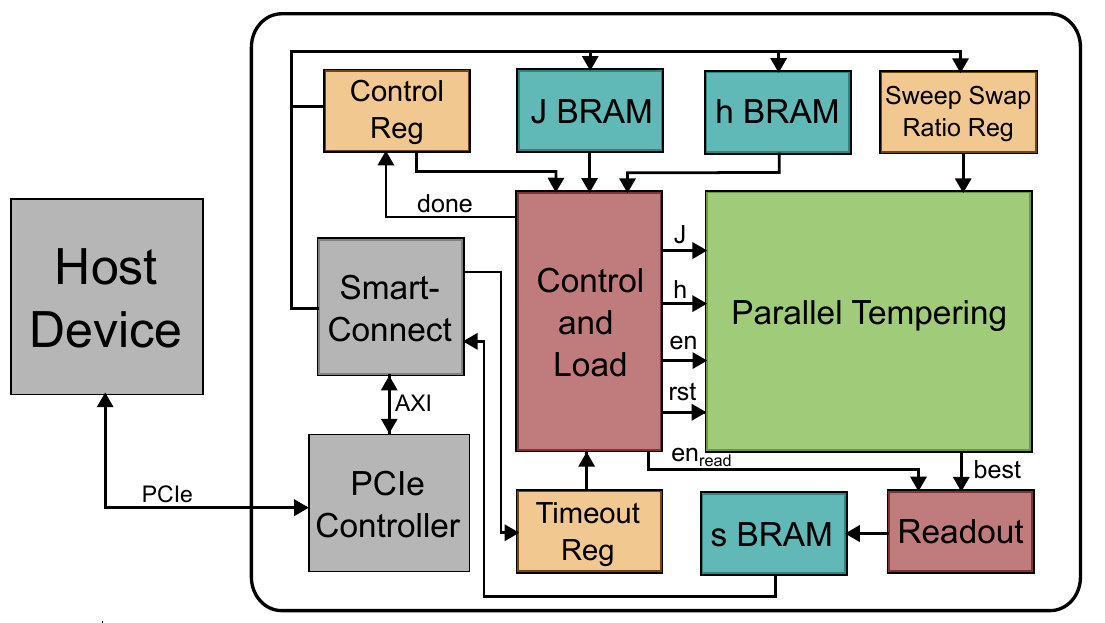}
\caption{Top-level architecture. The host device communicates with the FPGA over PCIe. The Control and Load module reads $\mathbf{J}$ and $\mathbf{h}$ from their respective BRAMs and distributes them to the Parallel Tempering core. When the run timer expires, the Readout module writes the best decoded states to the output BRAM for host retrieval. Control registers (start, load, sweep-to-swap ratio, timeout) are accessed through AXI GPIO.}
\label{fig:top}
\end{figure}

The coupling matrix $\mathbf{J}$ and bias vector $\mathbf{h}$ are written to dedicated BRAM cells using one of the two available BRAM access ports; the second port is reserved for the PT block. The decoded state vector is read from a third BRAM cell, which similarly uses one port for PT access and one for the communication block. Static control configurations, such as the sweep-to-swap ratio and run time, are written to AXI GPIO modules and typically remain constant across all test instances. A single AXI GPIO register is used for dynamic control signals: \texttt{load\_h}, \texttt{load\_j}, \texttt{start}, \texttt{rst\_best}, and the output \texttt{done} flag. The overall system architecture is shown in Figure~\ref{fig:top}.

\noindent{\textbf{Experimental Setup:}} FPGAs are connected to the host device via PCIe. The tests are performed in MATLAB, which interfaces with the PCIe AXI Controller for writing to memory and control registers. In a typical decoding experiment, the same channel is reused for multiple transmitted symbols: $\mathbf{J}$ is written once per channel realization, and $\mathbf{h}$ is written for every new symbol. The host device asserts the \texttt{load\_j} signal high to indicate that the coupling matrix must be reloaded before decoding. After writing the required memory contents and signaling the system to start, the host continuously polls the \texttt{done} flag until it is asserted, at which point the host reads the decoded state from its BRAM location.

\subsection{Modules}

This subsection describes all modules in the PT design hierarchy, proceeding from the lowest-level primitives to the top-level integration.

\subsubsection{P-bit}
\label{sec:pbit}
\begin{figure}[H]
\centering
\includegraphics[width=\textwidth]{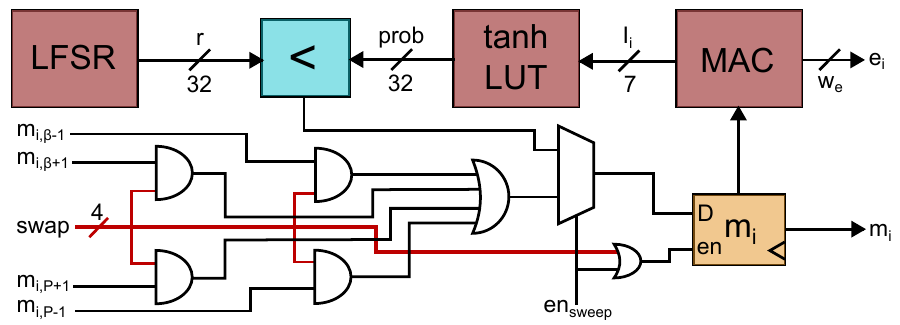}
\caption{p-bit block diagram. The MAC produces the 7-bit influence field $I_i$ and $w_e$-bit local energy $e_i$. The tanh LUT maps $I_i$ to a 32-bit probability, which is compared against the 32-bit LFSR output. The comparator result is muxed with four swap neighbor states through the 4-bit swap selector before reaching the state register.}
\label{fig:pbit}
\end{figure}

The probabilistic bit (p-bit) is the fundamental computational unit of the Ising machine (Figure~\ref{fig:pbit}). Each p-bit $i$ maintains a binary state $m_i \in \{0, 1\}$ (mapped to $\pm 1$ in the Ising formulation) and updates stochastically according to:
\begin{equation}
    m_i = \text{sgn}\bigl(\tanh(\beta I_i) - r_i\bigr)
\end{equation}
where $I_i$ is the local influence field, $\beta$ is the inverse temperature (which is precomputed into the weights), and $r_i \in [-1, 1]$ is a uniform pseudo-random number.

The p-bit module instantiates three sub-modules: the synapse (\texttt{weight}, \cref{sec:mac}), which computes the 7-bit influence field $I_i$ and the $w_e$-bit local energy $e_i$; the tanh lookup table (\texttt{tanh}, \cref{sec:tanh}), which maps $I_i$ to a 32-bit probability bias; and the LFSR (\cref{sec:lfsr}), which generates a 32-bit pseudo-random number $r_i$. The state update is performed in a single clocked register: on each enabled clock cycle, the p-bit compares the LFSR output against the tanh bias and latches the resulting binary state. As shown in Figure~\ref{fig:pbit}, the comparator output feeds the D input of the state register, which is clocked by \texttt{en\_sweep}.

In addition to the standard stochastic update, each p-bit supports state injection from up to four neighboring replicas for swap operations. The 4-bit \texttt{swap} input selects the source direction: \texttt{swap[0]} loads the state from the $\beta{-}1$ neighbor (hotter replica), \texttt{swap[1]} from $\beta{+}1$ (colder replica), \texttt{swap[2]} from $P{-}1$ (weaker constraint), and \texttt{swap[3]} from $P{+}1$ (stronger constraint). During normal Monte Carlo sweeps, all swap bits are deasserted and the p-bit updates stochastically. When a swap is accepted by the swap controller, the corresponding swap bit is asserted for one clock cycle, causing all p-bits in the target replica to simultaneously load the state vector from the source replica. In the 1D-PT configuration, only \texttt{swap[0:1]} are used; \texttt{swap[2:3]} are tied to zero.

\subsubsection{Synapse: MAC and Local Energy}
\label{sec:mac}
\begin{figure}[H]
\centering
\includegraphics[width=\textwidth]{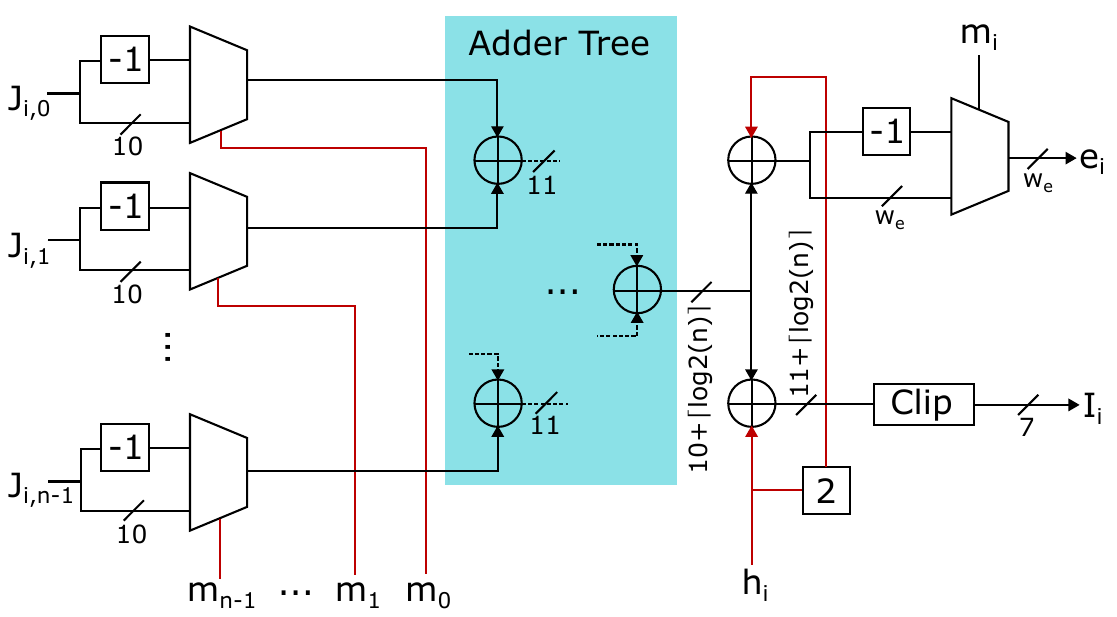}
\caption{Synapse: MAC and local energy. Each 10-bit weight $J_{ij}$ is conditionally negated by state $m_j$, then all operands are summed via a binary adder tree that grows by one bit per layer. The tree root at $(10 + \lceil\log_2 n\rceil)$ bits is split into two paths: one adds $h_i$ and clips to 7 bits for $I_i$, the other adds $2h_i$ and is conditionally negated by $m_i$ to produce the $w_e$-bit local energy $e_i$.}
\label{fig:mac}
\end{figure}

The synapse module computes two quantities for each p-bit $i$: the influence field $I_i$ and the local energy contribution $e_i$. The datapath is shown in Figure~\ref{fig:mac}.

\textbf{Influence field.} The influence field is the weighted sum of neighboring states plus the external bias:
\begin{equation}
    I_i = \sum_{j \in \mathcal{N}(i)} J_{ij}\, m_j + h_i
\end{equation}
Because p-bit states are binary ($m_j \in \{0,1\}$, mapped to $\pm 1$), the multiplication $J_{ij} \cdot m_j$ reduces to a sign-select operation: each 10-bit weight $J_{ij}$ is passed through unchanged if $m_j = 1$, or negated if $m_j = 0$. This eliminates the need for hardware multipliers entirely; each ``multiply'' is implemented as a conditional negation.

The sign-selected operands are summed using a parameterized binary adder tree. For a p-bit with $n$ neighbors, the tree has $\lceil \log_2 n \rceil$ layers and produces the result in a single combinational pass. As shown in Figure~\ref{fig:mac}, each layer grows the wire width by one bit to accommodate carry propagation: the inputs enter at 10 bits, the first adder layer produces 11-bit sums, and the root output is $(10 + \lceil\log_2 n\rceil)$ bits wide. When $n$ is not a power of two, the top layer of the tree passes through the excess operands without addition (a ``passthrough'' optimization that avoids wasting adder resources). The 10-bit bias $h_i$ is added to the tree output to produce the full-precision influence field at $(11 + \lceil\log_2 n\rceil)$ bits, which is then clipped to the $I_i$ output width (7 bits in Q3.3 format) to prevent overflow.

\textbf{Local energy.} The local energy contribution is computed simultaneously from the same adder tree output:
\begin{equation}
    e_i = -m_i \cdot (I_i + h_i)
    \label{eq:local_energy}
\end{equation}
where the factor of $1/2$ from the Ising Hamiltonian is absorbed into the final energy accumulation. The implementation takes the tree sum (before bias addition and before clipping, preserving the full $(10 + \lceil\log_2 n\rceil)$-bit precision) and adds twice the bias, then conditionally negates the result based on the p-bit's own state $m_i$. The result is the $w_e$-bit local energy output that feeds the energy accumulator. This is computed purely in combinational logic with no additional clock cycles.

\textbf{Fixed-point format.} All coupling weights $J_{ij}$ and biases $h_i$ use a 10-bit fixed-point representation (Q6.3). The inverse temperature $\beta$ is absorbed into the weights during preprocessing on the host, i.e., the CPU uploads $\beta_r J_{ij}$ and $\beta_r h_i$ directly, eliminating any on-chip multiplication by $\beta$.

\subsubsection{Tanh Lookup Table}
\label{sec:tanh}

The tanh activation function maps the clipped influence field $I_i$ to a 32-bit probability bias used for the stochastic state update, as shown in the top half of Figure~\ref{fig:tanh_lfsr}. The implementation uses a lookup table (LUT) with 32 entries, initialized from a memory file at synthesis time.

The module operates as follows. The sign bit $I_i[6]$ is separated from the magnitude $I_i[5{:}0]$. The magnitude is conditionally negated to produce a 6-bit unsigned absolute value. The lower 5 bits (3 fractional and 2 integer, covering the range $[0, 4)$ in steps of $0.125$) serve as the ROM address. If the MSB of the magnitude is set, indicating that $|I_i| \geq 4$ and the input has saturated beyond the LUT range, the output is clamped to the maximum value $\texttt{ONE} = 2^{32} - 1$, corresponding to $\tanh \to 1$. Otherwise, the ROM output provides the unsigned 32-bit magnitude of the bias.

Finally, a sign correction is applied. If the original input $I_i$ was negative, the bias magnitude is two's-complement negated to produce a signed 32-bit output. The result, combined with the LFSR output, determines the p-bit update, $m_i = 1$ if $r < \text{bias}$, and $m_i = 0$ otherwise.

\subsubsection{LFSR}
\label{sec:lfsr}

\begin{figure}[H]
\centering
\includegraphics[width=0.8\textwidth]{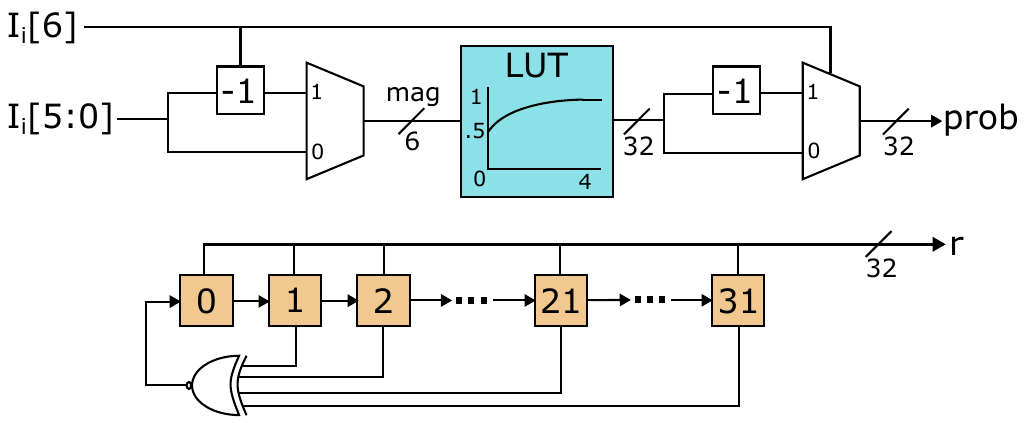}
\caption{Top: tanh lookup table. The 7-bit input $I_i$ is decomposed into sign and magnitude; the magnitude addresses a 32-entry ROM. The sign is reapplied to the ROM output to produce the 32-bit probability. Bottom: 32-bit Galois LFSR with feedback taps at bits 31, 21, 2, and 1.}
\label{fig:tanh_lfsr}
\end{figure}

Pseudo-random numbers for the stochastic p-bit updates and swap acceptance decisions are generated by Galois linear feedback shift registers (LFSRs), shown in the bottom half of Figure~\ref{fig:tanh_lfsr}. Each LFSR is a 32-bit register with feedback taps selected according to maximal-length polynomial specifications from Xilinx Application Note XAPP052. The specific taps for the 32-bit configuration are bits 31, 21, 2, and 1, producing a sequence with period $2^{32} - 1$.

Each p-bit and each swap controller instantiates its own LFSR with a unique randomly selected seed (parameterized at instantiation time), ensuring statistically independent random streams across all stochastic elements in the design. The LFSR advances by one step on every clock cycle, regardless of whether the p-bit is enabled, providing continuously refreshed randomness.

\subsubsection{Replica}
\label{sec:replica}

\begin{figure}[H]
\centering
\includegraphics[width=\textwidth]{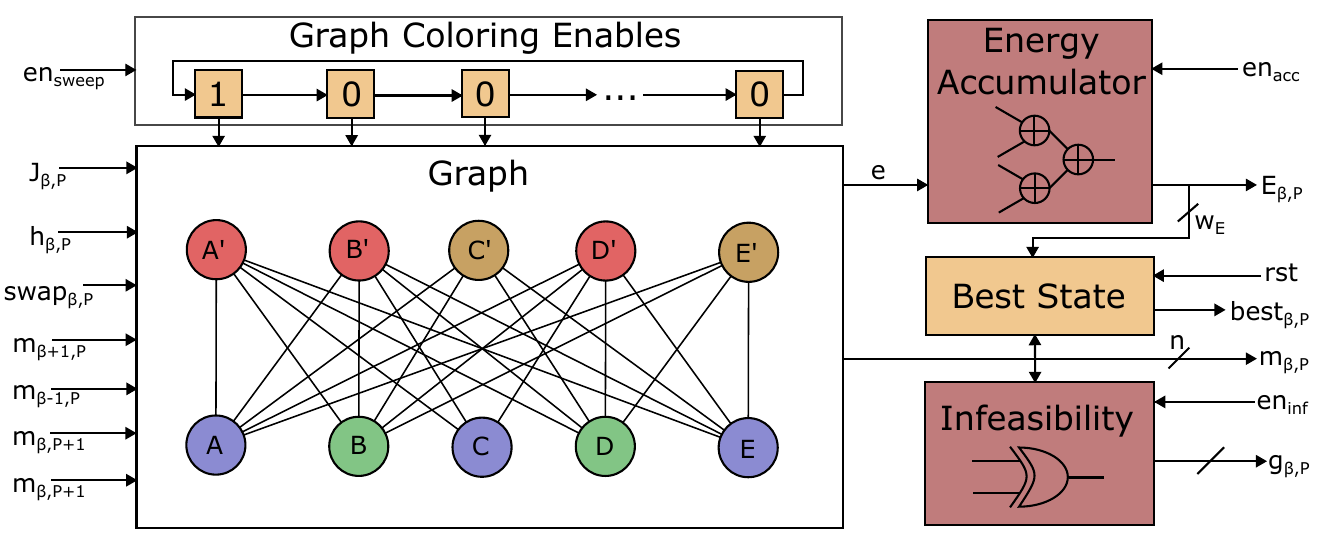}
\caption{Replica module. The graph coloring one-hot register at top enables one color group per cycle. All $N$ p-bits (here labeled A through E with copies A' through E') share inputs $\mathbf{J}_{\beta,P}$, $\mathbf{h}_{\beta,P}$, and the swap signals. The local energies feed the energy accumulator ($w_E$-bit output), and the state vector feeds both the best-state tracker and the infeasibility counter ($g_{\beta,P}$ output).}
\label{fig:replica}
\end{figure}

The replica module (Figure~\ref{fig:replica}) encapsulates a single replica: it instantiates all $N$ p-bits with their specific connectivity, along with the energy accumulator, infeasibility counter, and best-state tracker. The replica module is the primary unit of replication in the PT array; each position in the temperature/constraint grid is one instance of this module.

\textbf{Graph coloring.} To ensure that no two simultaneously updating p-bits share an edge (which would violate the sequential Gibbs sampling requirement), the $N$ p-bits are partitioned into $N_\text{color}$ groups by graph coloring. As shown at the top of Figure~\ref{fig:replica}, a rotating one-hot register cycles through the color groups on each clock cycle, enabling exactly one group per cycle. For the sparsified MIMO graphs used in this work, $N_\text{color} = 3$ colors suffice. One complete sweep through all color groups constitutes a single Monte Carlo sweep, requiring $N_\text{color}$ clock cycles.

\textbf{Energy accumulation.} At the end of each sweep phase (when signaled by \texttt{acc\_en}), the local energy contributions $e_i$ from all $N$ p-bits are captured into a shift register. The shift register feeds $d$ values per cycle (where $d$ is the stride parameter) into a pipelined reduction adder tree, which sums them into the total replica energy $E = \frac{1}{2}\sum_i e_i$. The accumulation requires $\lceil N/d \rceil$ clock cycles. The stride $d$ is a design parameter that trades accumulation latency against adder tree width; for the configurations in this work, $d = 32$ (1D-PT, $N = 128$) and $d = 8$ (2D-PT, $N = 32$). The output is the $w_E$-bit signed replica energy $E_{\beta,P}$.

\textbf{Infeasibility counter.} The infeasibility counter computes the number of copy-node disagreements in the sparsified graph. For a system with $N_\text{copies} = 2$ copies per original node, the $N$-bit state vector is split into two halves of $N/2$ bits. The XOR of the upper and lower halves produces a disagreement vector, and a population count yields the total number of disagreeing copy pairs as a $\lceil\log_2(N/2)\rceil$-bit unsigned integer. This value is used by the constraint swap controller (\cref{sec:swap_constraint}) to evaluate $P$-swap acceptance. The infeasibility is computed in a single clock cycle when signaled by \texttt{inf\_en}.

\textbf{Best energy tracking.} Each replica tracks its own best energy. After every $\beta$-swap phase, the current energy is compared against a stored best-energy register. If the current energy is less than or equal to the stored best, both the best energy and the corresponding $N$-bit state vector are updated. On reset, the best-energy register is initialized to the maximum positive value so that any valid energy will replace it. The best states are used during readout.

\subsubsection{Energy Accumulation}
\label{sec:accumulation}

\begin{figure}[H]
\centering
\includegraphics[width=0.8\textwidth]{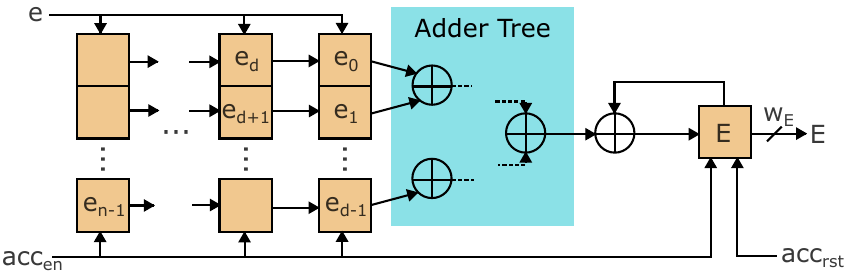}
\caption{Energy accumulator. The $N$-entry shift register captures all local energies on \texttt{acc\_en} and shifts by $d$ per cycle. The lowest $d$ entries feed a reduction adder tree whose output accumulates into the $w_E$-bit register $E$.}
\label{fig:energy}
\end{figure}

The energy accumulation module (Figure~\ref{fig:energy}) receives $d$ local energy values per clock cycle from the replica's shift register and reduces them to a running sum using a parameterized binary adder tree. The tree has $\lceil \log_2 d \rceil$ layers and handles non-power-of-two stride values via the same passthrough optimization used in the synapse adder tree (\cref{sec:mac}).

As shown in Figure~\ref{fig:energy}, the $N$ local energies are first captured into a shift register on the rising edge of \texttt{acc\_en}. On each subsequent cycle, the register shifts by $d$ positions, feeding the next window of $d$ values into the adder tree. The tree output is added to the $w_E$-bit accumulator register $E$. The accumulator is reset at the start of each energy phase (\texttt{acc\_rst} = rising edge \texttt{acc\_en}) and accumulates for $\lceil N/d \rceil$ cycles. The final value is the total replica energy $\beta_r E_r$, which is routed to the swap controller.

\subsubsection{Infeasibility}
\label{sec:infeasibility}

\begin{figure}[H]
\centering
\includegraphics[width=0.5\textwidth]{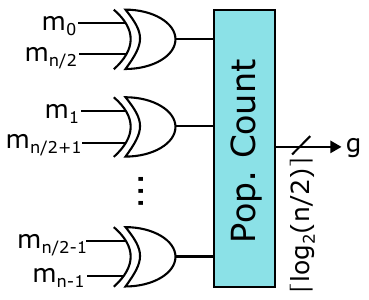}
\caption{Infeasibility counter. Copy pairs $(m_k, m_{k+N/2})$ are XORed and the results are summed by a population count to produce the $\lceil\log_2(N/2)\rceil$-bit disagreement count $g$.}
\label{fig:infeasibility}
\end{figure}

The infeasibility module (Figure~\ref{fig:infeasibility}) is described in \cref{sec:replica}. For a graph with $N$ physical nodes and $N_\text{copies} = 2$ copies per logical node, the module pairs corresponding nodes from the upper and lower halves of the state vector, XORs each pair, and returns the population count as a $\lceil\log_2(N/2)\rceil$-bit unsigned integer. A value of 0 indicates a fully feasible (all copies agree) state.

\subsubsection{Array}
\label{sec:array}

The array refers to the 2D grid of replica instantiations within the 2D-PT module. In 1D-PT, the array degenerates to a single column of $R$ replicas indexed by inverse temperature $\beta$. In 2D-PT, the array is an $R \times C$ grid where rows are indexed by $\beta$ (row 0 = hottest, row $R{-}1$ = coldest) and columns are indexed by constraint strength $P$ (column 0 = weakest, column $C{-}1$ = strongest). Each replica $(r, c)$ in the grid is connected to up to four neighbors for state exchange:
\begin{itemize}
    \item $\beta{-}1$: replica $(r{-}1, c)$, hotter temperature, same constraint
    \item $\beta{+}1$: replica $(r{+}1, c)$, colder temperature, same constraint
    \item $P{-}1$: replica $(r, c{-}1)$, same temperature, weaker constraint
    \item $P{+}1$: replica $(r, c{+}1)$, same temperature, stronger constraint
\end{itemize}
At the grid boundaries, the corresponding neighbor ports are tied to zero and the swap signals are deasserted, so edge and corner replicas simply have fewer active swap directions.

\textbf{Signal distribution.} The FSM control signals (sweep enable, accumulation enable, infeasibility enable, swap enables, direction flags, and readout signal) are distributed from the central FSM to all replicas through a pipelined shift register of depth $\lceil \log_2(R \times C) \rceil + 1$. Synthesis attributes (\texttt{equivalent\_register\_duplication} and \texttt{max\_fanout}) ensure that the tools replicate these pipeline registers as needed to meet timing across the large fanout.

\subsubsection{FSM}
\label{sec:fsm}

\begin{figure}[H]
\centering
\includegraphics[width=\textwidth]{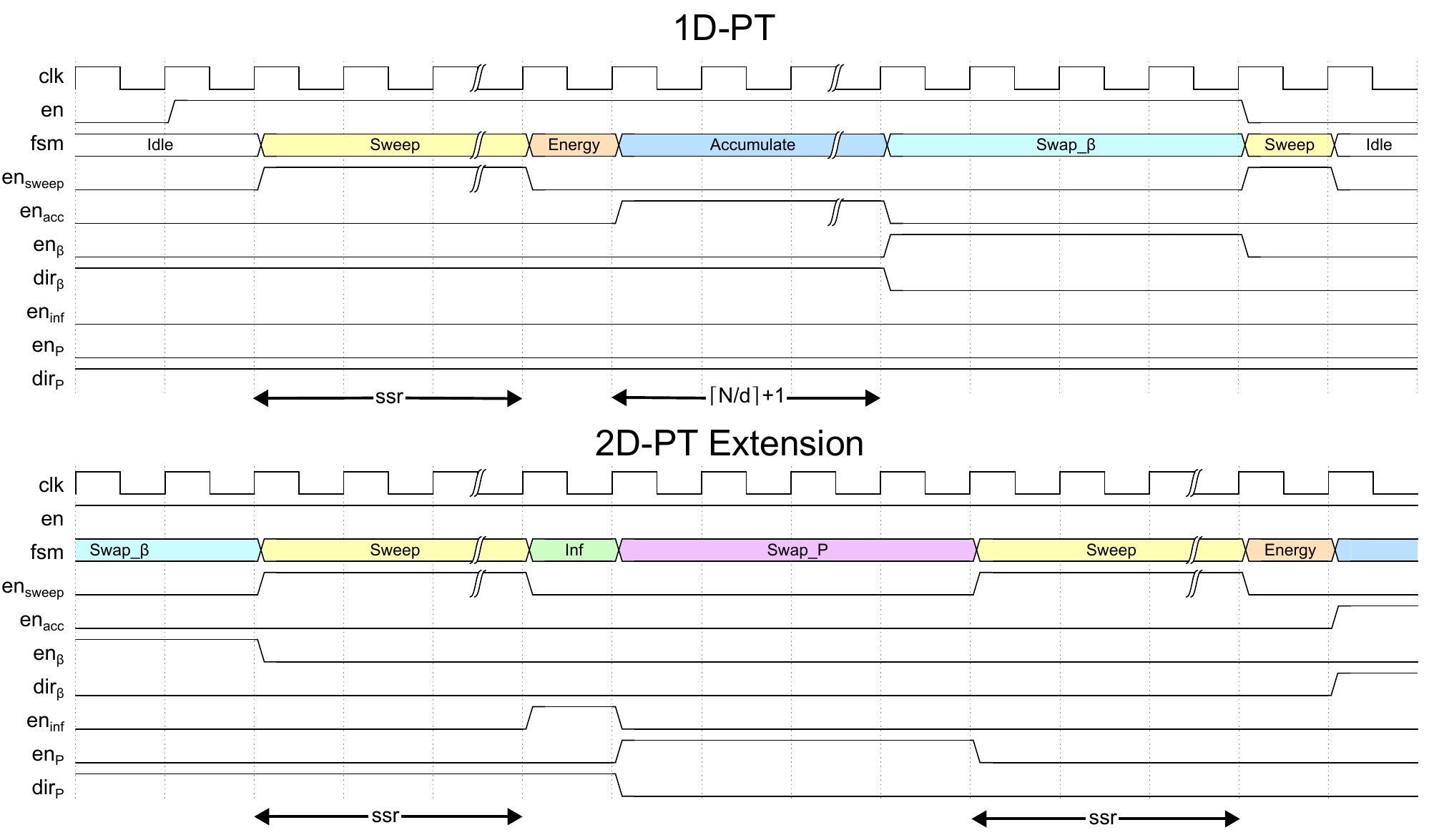}
\caption{FSM timing diagrams. Top: 1D-PT operation showing the Sweep$\to$Energy$\to$Acc$\to$Swap$_\beta$ loop with enable signals. The sweep phase lasts for the precomputed ``ssr'' = $N_\text{color} \times S$ cycles, and the accumulation phase lasts $\lceil N/d \rceil + 1$ cycles, where N is the number of nodes and d is the hyperparameter that controls the number of energies accumulated in a clock cycle. Bottom: 2D-PT extension showing the alternating $\beta$-swap and $P$-swap paths.}
\label{fig:fsm}
\end{figure}

The finite state machine (Figure~\ref{fig:fsm}) orchestrates the parallel tempering algorithm by cycling all replicas through a deterministic sequence of phases. The FSM has six states: \textsc{Idle}, \textsc{Sweep}, \textsc{Energy}, \textsc{Acc} (accumulation), \textsc{Infeas} (infeasibility calculation), and \textsc{Swap}.

\textbf{1D-PT operation.} In 1D-PT (single column, $C = 1$), the FSM executes the following loop:
\begin{equation*}
    \textsc{Sweep} \to \textsc{Energy} \to \textsc{Acc} \to \textsc{Swap}_{\ \beta} \to \textsc{Sweep} \to \cdots
\end{equation*}
The \textsc{Infeas} state is never entered because the constraint swap is not used. On each iteration, the $\beta$-swap direction alternates (even/odd pairing), controlled by the \texttt{dir\_b} flag which toggles after every swap phase. The top timing diagram in Figure~\ref{fig:fsm} shows this sequence, with the corresponding enable signals highlighted below the state trace.

\textbf{2D-PT operation.} In 2D-PT ($C > 1$), the FSM alternates between $\beta$-swaps and $P$-swaps on successive iterations. The alternation is governed by the condition $\texttt{dir\_b} \oplus \texttt{dir\_P}$: when the two direction flags differ, the FSM routes to the infeasibility/constraint-swap path instead of the energy/$\beta$-swap path. The resulting sequence is:
\begin{align*}
    &\textsc{Sweep} \to \textsc{Energy} \to \textsc{Acc} \to \textsc{Swap}_{\ \beta} \\
    \to\ &\textsc{Sweep} \to \textsc{Infeas} \to \textsc{Swap}_{\ P} \\
    \to\ &\textsc{Sweep} \to \textsc{Energy} \to \textsc{Acc} \to \textsc{Swap}_{\ \beta} \\
    \to\ &\textsc{Sweep} \to \textsc{Infeas} \to \textsc{Swap}_{\ P} \to \cdots
\end{align*}
This interleaving ensures that $\beta$-swaps and $P$-swaps occur at equal rates, each preceded by the appropriate metric computation (energy for $\beta$-swaps, infeasibility for $P$-swaps). The bottom timing diagram in Figure~\ref{fig:fsm} shows the 2D-PT extension, where the $P$-swap path appears between two $\beta$-swap iterations.

\textbf{Timing.} Each phase has a deterministic duration:
\begin{itemize}
    \item \textsc{Sweep}: $N_\text{sweep}$ cycles. The host precomputes $N_\text{sweep} = N_\text{color} \times S$ (where $S$ is the sweep-to-swap ratio) and writes this value to the sweep-swap ratio register. The FSM sweep timer counts up to this value. This corresponds to the ``ssr'' annotation in Figure~\ref{fig:fsm}.
    \item \textsc{Energy}: 1 cycle (propagates new states through the synapse to calculate current local energies).
    \item \textsc{Acc}: $\lceil N/d \rceil + 1$ cycles (shift-register-fed accumulation).
    \item \textsc{Infeas}: 1 cycle (single-cycle population count).
    \item \textsc{Swap}: 4 cycles (pipeline latency of the swap controller).
\end{itemize}
The total clock cycles per $\beta$-swap step are:
\begin{equation}
    C_{\text{step}, \beta} = N_\text{color} \cdot S + \lceil N/d \rceil + 6
    \label{eq:cstep_beta}
\end{equation}
and the total clock cycles per $P$-swap step are:
\begin{equation}
    C_{\text{step},P} = N_\text{color} \cdot S + 5
    \label{eq:cstep_p}
\end{equation}
(which skips the \textsc{Energy} and \textsc{Acc} phases). A full step is considered as the time to do an attempted $\beta$-swap and a $P$-swap, meaning the total step time for the PT system is $C_\text{step} = C_{\text{step},\beta} + C_{\text{step},P}$. Since the constraint swap does not occur in 1D-PT that value is set to 0.

\subsubsection{Swap Beta}
\label{sec:swap_beta}

The $\beta$-swap controller evaluates the Metropolis acceptance criterion for exchanging configurations between adjacent replicas along the temperature axis. Each swap controller instance manages three consecutive replicas and produces two independent swap decisions (for the two adjacent pairs), with the active pair selected by the alternating direction flag \texttt{dir\_b}.

\textbf{Log-probability computation.} The Metropolis acceptance probability for swapping replicas $a$ and $b$ is $p_\text{swap} = \min(1, e^{\Delta})$, where $\Delta = (\beta_b - \beta_a)(E_b - E_a)$. Since each replica module provides only the temperature-scaled energy $\beta_r E_r$, the log-probability argument is expanded as:
\begin{equation}
    \Delta = \mu_0 \cdot (\beta_a E_a) + \mu_1 \cdot (\beta_b E_b)
    \label{eq:logprob_beta}
\end{equation}
where $\mu_0 = 1 - \beta_b / \beta_a$ and $\mu_1 = 1 - \beta_a / \beta_b$ are precomputed scaling factors. The $\mu$ values are stored as fixed-point integers with $\mu_\text{decimal} = 3$ fractional bits and are provided as compile-time parameters to each swap controller instance. Each controller stores four $\mu$ values: two for the even-indexed pair and two for the odd-indexed pair.

\textbf{Exponential approximation.} The exponential $e^{\Delta}$ is computed using a base-2 approximation optimized for fixed-point arithmetic:
\begin{equation}
    e^x = 2^{x \log_2 e} \approx 2^{\lfloor \frac{23}{16} x \rfloor} \left(1 + \frac{23}{16}x - \left\lfloor \frac{23}{16}x \right\rfloor \right)
    \label{eq:exp_approx}
\end{equation}
The multiplication by $23/16$ is implemented without a hardware multiplier using bit-shifts and additions. The result is decomposed into a 5-bit exponent and a 5-bit mantissa, yielding a compact pseudo-floating-point representation.

The \texttt{exp} module also handles domain boundaries: if the input $\Delta$ is positive, the exponential exceeds 1 and the swap is automatically accepted (overflow flag); if $\Delta$ is below the representable negative range, the exponential is effectively zero and the swap is rejected (underflow flag).

\textbf{Random number conversion.} To compare the exponential output against a uniform random number, the 32-bit LFSR output must be converted to the same pseudo-floating-point format. A submodule called \texttt{float\_Mplus5} implements this conversion using a priority encoder (leading-one detector): a \texttt{priority casez} statement identifies the position of the most significant set bit in the 32-bit input, producing a 5-bit exponent. The mantissa is then extracted as the $M$ bits immediately below the leading one. This allows a direct comparison of exponent and mantissa fields without requiring a full-precision division or normalization.

\textbf{Acceptance logic.} The swap is accepted if the exponential output is greater than or equal to the random number, first comparing the exponents and the mantissa in case of an exponent tie. The acceptance decision propagates through a 3-stage pipeline to align with the log-probability and exponential computation latencies. The final swap signal is gated by the direction flag: \texttt{swap[0]} fires on even cycles (pairing replicas 0--1) and \texttt{swap[1]} on odd cycles (pairing replicas 1--2).\\

\subsubsection{Swap Constraint}
\label{sec:swap_constraint}

The $P$-swap (constraint swap) controller is structurally identical to the $\beta$-swap controller, but operates on infeasibility counts rather than energies. For two adjacent replicas along the constraint axis (columns $c$ and $c{+}1$ at the same row $r$), the log-probability is:
\begin{equation}
    \Delta = \mu \cdot (g_{c+1} - g_c)
    \label{eq:logprob_constraint}
\end{equation}
where $g_c$ is the infeasibility count (number of disagreeing copy pairs) for the replica at column $c$, and $\mu$ is a precomputed scaling factor related to the constraint-strength difference $\Delta P$ between the two columns. The same exponential approximation (\cref{eq:exp_approx}), floating-point conversion, and acceptance pipeline are used.

The key behavioral difference is that constraint swaps encourage the migration of feasible states (low $g$) from weak-constraint columns (where mixing is fast) to strong-constraint columns (where constraints are strictly enforced but mixing is slow). This is the mechanism by which 2D-PT eliminates the need for manual $P$-tuning.

\subsubsection{PT}
\label{sec:pt}

\begin{figure}[H]
\centering
\includegraphics[width=\textwidth]{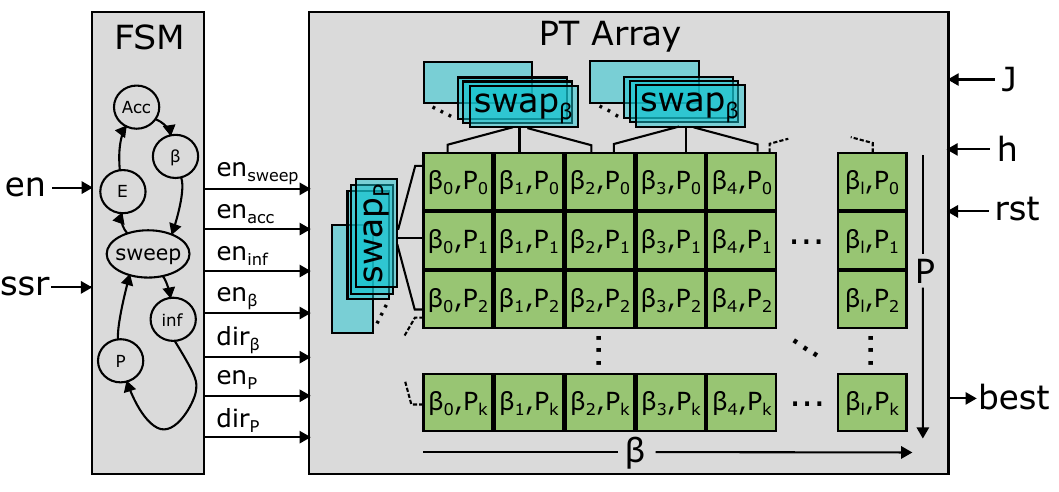}
\caption{PT module. The FSM (left) drives control signals to the $R \times C$ replica array. Swap-$\beta$ controllers operate along columns and swap-$P$ controllers along rows. The best states from the rightmost $P$-column are read out. External inputs $\mathbf{J}$, $\mathbf{h}$, and \texttt{rst} are distributed to all replicas; the sweep-to-swap ratio (ssr) configures the FSM.}
\label{fig:pt}
\end{figure}

The top-level PT module (Figure~\ref{fig:pt}) integrates the FSM, the replica array, all swap controllers, and the readout shift register.

\textbf{Replica instantiation.} The module instantiates $R \times C$ replica modules, each parameterized with a unique LFSR seed offset to ensure independent random streams. The state, energy, and infeasibility outputs of each replica are collected into 2D arrays indexed by $[r][c]$. As shown in Figure~\ref{fig:pt}, the array is laid out with $\beta$ increasing downward (columns) and $P$ increasing rightward (rows). \textit{Note: Figure~\ref{fig:pt} illustrates a transposed $\beta,P$ array, however this does not affect the systems function.}

\textbf{Swap controller instantiation.} Along the $\beta$-axis, $\lfloor R/2 \rfloor \times C$ swap-beta controllers are instantiated, each managing a group of three consecutive replicas within a single column. Along the $P$-axis, $R \times \lfloor C/2 \rfloor$ swap-constraint controllers are instantiated, each managing three consecutive replicas within a single row. The $\mu$ parameters for each controller are precomputed from the $\beta$ and $P$ schedules and provided as compile-time constants. In Figure~\ref{fig:pt}, the swap controllers are shown along the top ($\text{swap}_\beta$) and left ($\text{swap}_P$) edges of the array.

\textbf{Readout.} When the system timer expires and \texttt{clk\_en} is deasserted, the FSM enters the readout phase. For 1D-PT, the best states of all $R$ replicas are packed into a shift register and shifted out 32 bits at a time through the BRAM interface. For 2D-PT, the best states of all $R$ replicas in the last column (strongest constraint, $c = C{-}1$) are read out, as this column enforces near-perfect copy agreement and holds the most feasible solutions. In Figure~\ref{fig:pt}, the readout path exits from the rightmost column. The host selects the minimum-energy state among the read-out replicas after mapping back to the original all-to-all problem via majority voting.

\subsubsection{Control and Load}
\label{sec:control}

The control and load module manages the initialization of all replica weights and the system-level control flow. It implements two daisy-chain shift registers, one for coupling weights $J$ and one for biases $h$, that are loaded sequentially from BRAM.

\textbf{Loading mechanism.} When the host asserts \texttt{load\_j} (or \texttt{load\_h}), the module begins reading 32-bit words from the corresponding BRAM at incrementing addresses. Each word is shifted into the head of the daisy-chain register, pushing all previous entries forward by one position. After $N_\text{replicas} \times N_\text{edges}$ words (for $J$) or $N_\text{replicas} \times N_\text{nodes}$ words (for $h$), the loading is complete and the shift register contents are mapped to the individual replica weight arrays via generate-block indexing. This approach allows all replicas to be loaded through a single BRAM port, at the cost of a linear loading time that scales with the total number of weights. Future implementations could reduce the loading time of the 2D-PT implementation by loading a $P$-column with information and copying those values to the remaining $P$-columns. This would require loading edges between copy nodes separately from the rest of the edges, as these are the only weights that differ between replicas in a $\beta$-row.

\textbf{System control.} The module implements a countdown timer that controls the duration of the PT run. When the host asserts \texttt{start}, the timer is initialized from the \texttt{IM\_Timer} register and begins counting. While the timer is active, the \texttt{clk\_en} signal is asserted, driving the PT FSM. When the timer expires, \texttt{clk\_en} is deasserted, causing the FSM to return to \textsc{Idle} and triggering the readout phase. The \texttt{done} flag is asserted after a short pipeline delay (to allow the readout to complete) and remains high until the next \texttt{start} pulse.

A separate \texttt{rst\_best} pulse, triggered by the host, resets all best-energy registers to their maximum positive value, allowing a fresh best-energy search without reloading weights.

\subsubsection{Readout}
\label{sec:readout}

The readout module transfers the best-state shift register contents to the output BRAM for host retrieval. When the readout signal (\texttt{s\_valid}) is first asserted (indicating that the 2D-PT module has latched the best states into its readout register), the module begins writing 32-bit words to sequential BRAM addresses. On each subsequent cycle, the 2D-PT readout register shifts left by 32 bits, presenting the next word at the output. The process continues for $\lceil (R \times N) / 32 \rceil$ words (1D-PT reads all replicas; 2D-PT reads the strongest-constraint column), after which the write-enable is deasserted and the \texttt{finished} flag is set.

\subsubsection{Hierarchy}
\label{sec:hierarchy}

The complete design hierarchy is summarized below, from top to bottom:

\begin{enumerate}
    \item \textbf{IsingMachine} (top level): PCIe interface wrapper, BRAM instantiation, clock domain crossing.
    \begin{enumerate}
        \item \textbf{IsingStart} (control and load): BRAM-to-register daisy-chain loader, run timer, system control signals.
        \item \textbf{PT2D} (parallel tempering core): FSM, replica array, swap controllers, readout shift register.
        \begin{enumerate}
            \item \textbf{PT\_FSM}: Six-state finite state machine orchestrating sweep/energy/accumulation/infeasibility/swap phases.
            \item \textbf{IsingGraph} $\times (R \times C)$: Individual replica modules, each containing:
            \begin{enumerate}
                \item \textbf{Graph}: $N$ p-bit instances with problem-specific connectivity and graph coloring.
                \begin{enumerate}
                    \item \textbf{pbit} $\times N$: Stochastic update unit.
                    \item \textbf{weight} (synapse/MAC): Adder-tree MAC, local energy.
                    \item \textbf{tanh}: ROM lookup table.
                    \item \textbf{LFSR\_n}: 32-bit Galois LFSR.
                \end{enumerate}
                \item \textbf{energy}: Pipelined reduction adder tree for energy accumulation.
                \item \textbf{infeasibility}: XOR + population count for copy disagreement.
            \end{enumerate}
            \item \textbf{SwapBeta} $\times (\lfloor R/2 \rfloor \times C)$: $\beta$-direction Metropolis swap controllers.
            \begin{enumerate}
                \item \textbf{exp}: Fixed-point exponential approximation.
                \item \textbf{float\_Mplus5}: Leading-one detector for LFSR-to-float conversion.
                \item \textbf{LFSR\_n}: Independent random stream for acceptance.
            \end{enumerate}
            \item \textbf{SwapConstraint} $\times (R \times \lfloor C/2 \rfloor)$: $P$-direction Metropolis swap controllers (same sub-modules as SwapBeta).
        \end{enumerate}
        \item \textbf{IsingReadout}: BRAM writer for decoded state output.
    \end{enumerate}
\end{enumerate}

\subsection{Design Parameters}

\Cref{tab:params} summarizes the key compile-time parameters that configure the design for a specific problem instance.

\begin{table}[H]
\centering
\caption{Key design parameters}
\label{tab:params}
\begin{tabular}{@{}lll@{}}
\toprule
Parameter & Symbol & Description \\
\midrule
\texttt{num\_nodes} & $N$ & Number of physical p-bits per replica \\
\texttt{num\_edges} & $|\mathcal{E}|$ & Number of edges in the sparsified graph \\
\texttt{rows} & $R$ & Number of $\beta$ (temperature) levels \\
\texttt{columns} & $C$ & Number of $P$ (constraint) levels \\
\texttt{num\_copies} & $N_\text{copies}$ & Copies per logical node (2 for all experiments) \\
\texttt{clk\_colors} & $N_\text{color}$ & Graph chromatic number for update scheduling \\
\texttt{stride} & $d$ & Parallel inputs per cycle to energy adder tree \\
\texttt{j\_bit\_width} & $w_J$ & Bit width of coupling weights (10 bits) \\
\texttt{h\_bit\_width} & $w_h$ & Bit width of bias values (10 bits) \\
\texttt{i\_bit\_width} & $w_I$ & Bit width of influence field (7 bits) \\
\texttt{max\_depth} & $\|\mathcal{E}\|_\infty$ & Maximum number of nodes connections \\
\texttt{e\_bit\_width} & $w_e$ & Bit width of local energy ($\log_2(w_J \cdot \|\mathcal{E}\|_\infty + 2w_h)$) \\
\texttt{E\_bit\_width} & $w_E$ & Bit width of replica energy ($\log_2N w_e$) \\
\texttt{mu\_decimal\_bits} &  & Fractional bits in swap scaling factors (3 bits) \\
\texttt{rng\_bit\_width} &  & LFSR output width (32 bits) \\
\bottomrule

\end{tabular}
\end{table}

\rsection{FPGA Resources Utilization}

\textbf{FPGA:} These implementations target the \textbf{Alveo U250} (part: \texttt{xcu250\--figd2104\--2L\--e}) accelerator card. All synthesis was done on Vivado ML edition 2024.1.

\textbf{Host interface:}
The host CPU communicates via the Xilinx \texttt{AMD DMA/Bridge Subsystem for PCI Express} v4.1 IP managed via the Mathworks' \texttt{PCIe AXI Manager} . This interface utilizes a PCIe Gen 3 with a link speed of 2.5 GT/s across an x8 lane width. The AXI interface operates at a clock frequency ($f_{\text{PCIe}}$) of 125 MHz and a data width of 128 bits. 

\begin{figure}[t]
\centering
\includegraphics[width=0.68\textwidth]{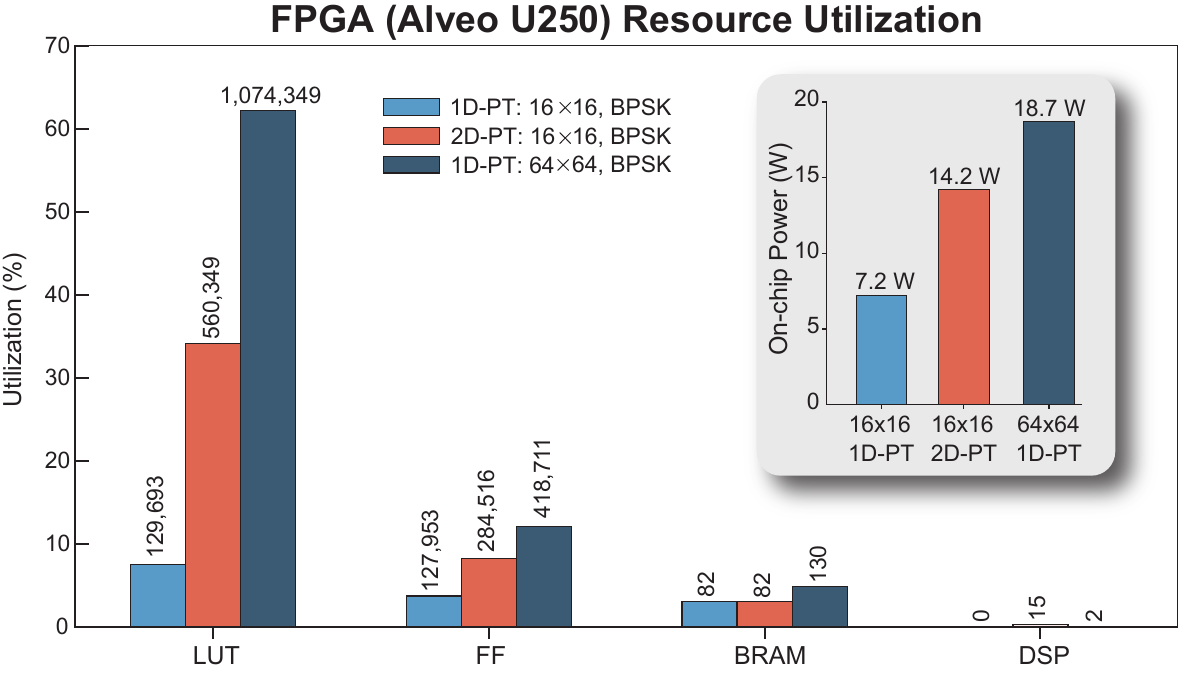}
    \caption{FPGA resource utilization and power consumption comparison for three Alveo U250 (part: \texttt{xcu250\--figd2104\--2L\--e}) hardware configurations. The main bar chart illustrates the percentage and raw counts of lookup tables (LUT), Flip-Flops (FF), Block RAM (BRAM), and DSP slices utilized for the 1D-PT ($16 \times 16$ BPSK), 2D-PT ($16 \times 16$ BPSK), and 1D-PT ($64 \times 64$ BPSK) implementations. The inset bar chart highlights the total on-chip power consumption, scaling from 7.2 W (28.8$^\circ$C) for the $16 \times 16$ 1D-PT design to 18.7 W (34.8$^\circ$C) for the $64 \times 64$ configuration. For $16 \times 16$ 2D-PT design, it takes 14.22 W (32.4$^\circ$C).   All designs achieved timing closure at operating frequencies between 70 and 80 MHz with positive Worst Negative Slack (WNS).}
\label{fig:fpga_resource}
\end{figure}

\textbf{Resource utilization:}
Resource utilization on the Alveo U250 across the evaluated MIMO implementations is detailed in Figure~\ref{fig:fpga_resource}. Transitioning the MIMO 16 design from a 1D-PT to a 2D-PT architecture increases lookup table (LUT) utilization from 7.51\% (129,693 slices) to 34.16\% (560,349 slices). This shift also results in an increased power usage, rising from 7.216 W (at a junction temperature of 28.8$^\circ$C) for the 1D model to 14.223 W (at a junction temperature of 32.4$^\circ$C) for the 2D configuration, while both sustain an 80 MHz clock frequency (WNS = 0.809 ns and 0.399 ns, respectively). The larger problem instance, a $64 \times 64$ 1D-PT implementation, requires nearly 1.1 million LUTs (62\% of capacity) and the highest power draw of 18.7 W (at a junction temperature of 34.8$^\circ$C), at a clock frequency of 70 MHz (WNS = 0.031~ns). Notably, digital signal processing (DSP) and block RAM (BRAM) utilization remain exceptionally low across all configuration types shown in Figure~\ref{fig:fpga_resource}. All implementations reported congestion less than level 5.

\rsection{Throughput and Energy-per-bit Analysis}
 
Tables~\ref{tab:S2_energy_per_bit_BER1e-3} and \ref{tab:S3_energy_per_bit_lowestBER} compare the throughput, energy per bit ($E_b$), and energy-delay product (EDP) per bit of the MIMO solver across FPGA (Alveo U250) and ASIC (ASAP7) platforms. The FPGA instance time, $t_{\text{inst}}$, is calculated using measured step durations ($t_{\text{step,meas}}$) and explicitly includes the PCIe load/read-out latency ($t_{\text{OH}}$), $t_\text{inst} = t_\text{step,meas} N_\text{steps} + t_{OH}$. In contrast, the ASIC projections rely on theoretical step times ($t_\text{step,theory}$) operating entirely on-chip and exclude host communication overhead, $t_\text{inst} = t_\text{step,theory} N_\text{steps}$. Theoretical step times are calculated according to the cycles required for a step, defined in Eq.~\ref{eq:cstep_beta} and~\ref{eq:cstep_p}, multiplied by the projected clock period. 

For a target BER of $10^{-3}$ (Table~\ref{tab:S2_energy_per_bit_BER1e-3}), the $64\!\times\!64$ 1D-PT FPGA configuration requires $N_{\text{steps}}=300$, resulting in an $E_b$ of $864\,\mu\text{J/bit}$ and an EDP of $2556\,\text{nJ-s/bit}$ \cite{han2022low, chi2025multilinear, xiang2025area, castaneda2022283}. Shrinking the problem to $16\!\times\!16$ 1D-PT reduces the energy cost to $364\,\mu\text{J/bit}$ and drops the EDP to $294\,\text{nJ-s/bit}$. Applying the 2D-PT architecture to the $16\!\times\!16$ problem significantly accelerates algorithmic convergence, requiring only $15$ steps. However, the compute time becomes so brief that the fixed $1.05\,\text{ms}$ PCIe overhead completely dominates $t_{\text{inst}}$. As a result, the FPGA energy efficiency degrades to $960\,\mu\text{J/bit}$ (with an EDP of $1037\,\text{nJ-s/bit}$), demonstrating how host-FPGA communication bottlenecks can entirely mask the algorithmic advantages of 2D-PT on this platform.

On dedicated silicon, this I/O bottleneck can be reduced, allowing the 2D-PT architecture to demonstrate its throughput advantage. The $16\!\times\!16$ 2D-PT ASIC achieves an exceptional $174\,\text{nJ/bit}$, a $\sim$$5{,}500\times$ improvement in energy over its FPGA counterpart. Furthermore, because both latency and energy consumption collapse without the communication bottleneck, its EDP plummets to $3.89\times 10^{-3}\,\text{nJ-s/bit}$, an improvement of approximately $2.6\times 10^5$ over the FPGA. However, as shown in Table~\ref{tab:S3_energy_per_bit_lowestBER}, driving the solver to its absolute lowest measured BER exposes a steep energy-accuracy trade-off. Reaching the minimum BER demands up to $10{,}000$ steps for the $64\!\times\!64$ problem, pushing $t_{\text{inst}}$ to $46.2\,\text{ms}$ and a projected $30.1$ms on the FPGA ($6.24\times 10^5\,\text{nJ-s/bit}$ EDP) and ASIC ($4045\,\text{nJ-s/bit}$ EDP), respectively. Finally, it should be noted that the ASAP7 entries are predictive projections (estimated at a $50\%$ switching activity factor); a taped-out variant in a commercial node would yield tighter area metrics while preserving these broad $E_b$ and EDP trends.

Therefore, an LTE slot-time budget of $8.4\!\times\!10^{6}$ instances/s \cite{singh2022ising} requires \emph{coarse-grained parallelism across many chips}, not pipelining within a single chip. Using our most efficient point ($16\!\times\!16$ 2D-PT ASIC at $124\,\text{mW}$ per chip, $5.76\,\text{mm}^2$), $\sim$$188$ chips run in parallel would meet the requirement. 

\clearpage

\begin{table*}[t]
  \centering
  \caption{Energy per bit and energy-delay product (EDP) per bit at a target BER of $10^{-3}$. The FPGA instance time is
  $t_{\text{inst}} = N_{\text{steps}}\cdot t_{\text{step,meas}} + t_{\text{OH}}$ (the measured wall-clock $t_{\text{step}}$ is used because it captures FSM overhead beyond the theoretical $C_{\text{step}}/f_{\text{clk}}$); ASIC entries use $t_{\text{step,theory}}$ since silicon measurements are unavailable, and $t_{\text{OH}}=0$ because no PCIe path is involved. }
  \label{tab:S2_energy_per_bit_BER1e-3}
  \renewcommand{\arraystretch}{1.2}
  \setlength{\tabcolsep}{3.5pt}
  \footnotesize
  \resizebox{\textwidth}{!}{%
  \begin{tabular}{l l c c c c c c c c c c}
    \toprule
    \textbf{Platform} & \textbf{Problem} & \textbf{Power} & \textbf{Clock} &
    $\boldsymbol{t_{\text{step,theory}}}$ & $\boldsymbol{t_{\text{step,meas}}}$ &
    $\boldsymbol{N_{\text{steps}}}$ & $\boldsymbol{t_{\text{OH}}}$ &
    $\boldsymbol{t_{\text{inst}}}$ & \textbf{Throughput} &
    \textbf{Energy/bit} & \textbf{EDP/bit} \\
    \midrule
    FPGA & $64\!\times\!64$ 1D-PT & $18.7\,\text{W}$   & $70\MHz$  & $4.43\mus$  & $4.46\mus$ & $300$ & $1.62\ms$  & $2.96\ms$  & $21.6\kbps$ & $864\uJpb$  & $2556\,\text{nJ$\cdot$s/bit}$ \\
    FPGA & $16\!\times\!16$ 1D-PT & $7.22\,\text{W}$   & $80\MHz$  & $1.06\mus$  & $1.07\mus$ & $100$ & $0.70\ms$  & $0.807\ms$ & $19.8\kbps$ & $364\uJpb$  & $294\,\text{nJ$\cdot$s/bit}$  \\
    FPGA & $16\!\times\!16$ 2D-PT & $14.22\,\text{W}$  & $80\MHz$  & $2.06\mus$  & $2.11\mus$ & $15$  & $1.05\ms$  & $1.08\ms$  & $14.8\kbps$ & $960\uJpb$  & $1037\,\text{nJ$\cdot$s/bit}$ \\
    \midrule
    ASIC (Proj.) & $64\!\times\!64$ 1D-PT & $285.8\,\text{mW}$ & $103\MHz$ & $3.01\mus$  & ---        & $300$ & ---        & $0.903\ms$ & $70.9\kbps$ & $4.03\uJpb$ & $3.64\,\text{nJ$\cdot$s/bit}$ \\
    ASIC (Proj.) & $16\!\times\!16$ 1D-PT & $19.6\,\text{mW}$  & $157\MHz$ & $0.541\mus$ & ---        & $100$ & ---        & $54.1\mus$ & $296\kbps$  & $66.3\,\text{nJ/bit}$ & $3.59\times 10^{-3}\,\text{nJ$\cdot$s/bit}$ \\
    ASIC (Proj.) & $16\!\times\!16$ 2D-PT & $124\,\text{mW}$   & $111\MHz$ & $1.49\mus$  & ---        & $15$  & ---        & $22.4\mus$ & $714\kbps$  & $174\,\text{nJ/bit}$  & $3.89\times 10^{-3}\,\text{nJ$\cdot$s/bit}$ \\
    \bottomrule
  \end{tabular}%
  }
\\[2pt]
\begin{flushleft}\footnotesize
\begin{itemize}
    \setlength{\itemsep}{0pt}
    \item $t_{\text{step,theory}}=C_{\text{step}}/f_{\text{clk}}$ where $C_{\text{step}}=C_{\text{step},\beta}$ for 1D-PT and $C_{\text{step}}=C_{\text{step},\beta}+C_{\text{step},P}$ for 2D-PT. Here, $C_{\text{step},\beta}=N_{\text{color}} \cdot S+\lceil N/d\rceil+6$ and $C_{\text{step},P}=N_{\text{color}} \cdot S+5$.
    \item $N_{\text{color}}=3$, $\lceil N/d\rceil=4$, Swap ratio, $S=100$ for $64\times64$ and $S=25$ for $16\times16$.
    \item $t_{\text{step,meas}}$ is the slope and $t_{\text{OH}}$ is the intercept of $t_{\text{inst}}$ versus $N_{\text{steps}}$ measured on the Alveo U250.
    \item Throughput is $N_{\text{bits}}/t_{\text{inst}}$, energy per bit is $\text{Power}/\text{Throughput}$, and EDP/bit is $\text{Energy/bit}\times t_{\text{inst}}$.
\end{itemize}
\end{flushleft}
\end{table*}

\begin{table*}[t]
  \centering
  \caption{Energy per bit and EDP per bit at the lowest measured BER for each
  configuration. All formulas, definitions, and footnotes are identical to
  Table~\ref{tab:S2_energy_per_bit_BER1e-3}; only $N_{\text{steps}}$ and the
  derived quantities ($t_{\text{inst}}$, Throughput, Energy/bit, EDP/bit) change.}
  \label{tab:S3_energy_per_bit_lowestBER}
  \renewcommand{\arraystretch}{1.2}
  \setlength{\tabcolsep}{3.5pt}
  \footnotesize
  \resizebox{\textwidth}{!}{%
  \begin{tabular}{l l c c c c c c c c c c}
    \toprule
    \textbf{Platform} & \textbf{Problem} & \textbf{Power} & \textbf{Clock} &
    $\boldsymbol{t_{\text{step,theory}}}$ & $\boldsymbol{t_{\text{step,meas}}}$ &
    $\boldsymbol{N_{\text{steps}}}$ & $\boldsymbol{t_{\text{OH}}}$ &
    $\boldsymbol{t_{\text{inst}}}$ & \textbf{Throughput} &
    \textbf{Energy/bit} & \textbf{EDP/bit} \\
    \midrule
    FPGA & $64\!\times\!64$ 1D-PT & $18.7\,\text{W}$   & $70\MHz$  & $4.43\mus$  & $4.46\mus$ & $10{,}000$ & $1.62\ms$ & $46.2\ms$  & $1.39\kbps$ & $13.5\mJpb$ & $6.24\times 10^{5}\,\text{nJ$\cdot$s/bit}$ \\
    FPGA & $16\!\times\!16$ 1D-PT & $7.22\,\text{W}$   & $80\MHz$  & $1.06\mus$  & $1.07\mus$ & $500$      & $0.70\ms$ & $1.24\ms$  & $13.0\kbps$ & $560\uJpb$  & $694\,\text{nJ$\cdot$s/bit}$  \\
    FPGA & $16\!\times\!16$ 2D-PT & $14.22\,\text{W}$  & $80\MHz$  & $2.06\mus$  & $2.11\mus$ & $50$       & $1.05\ms$ & $1.16\ms$  & $13.8\kbps$ & $1.03\mJpb$ & $1196\,\text{nJ$\cdot$s/bit}$ \\
    \midrule
    ASIC (Proj.) & $64\!\times\!64$ 1D-PT & $285.8\,\text{mW}$ & $103\MHz$ & $3.01\mus$  & ---        & $10{,}000$ & ---       & $30.1\ms$  & $2.13\kbps$ & $134\uJpb$  & $4045\,\text{nJ$\cdot$s/bit}$ \\
    ASIC (Proj.) & $16\!\times\!16$ 1D-PT & $19.6\,\text{mW}$  & $157\MHz$ & $0.541\mus$ & ---        & $500$      & ---       & $271\mus$  & $59.0\kbps$ & $332\,\text{nJ/bit}$  & $9.00\times 10^{-2}\,\text{nJ$\cdot$s/bit}$ \\
    ASIC (Proj.) & $16\!\times\!16$ 2D-PT & $124\,\text{mW}$   & $111\MHz$ & $1.49\mus$  & ---        & $50$       & ---       & $74.5\mus$ & $215\kbps$  & $577\,\text{nJ/bit}$  & $4.30\times 10^{-2}\,\text{nJ$\cdot$s/bit}$ \\
    \bottomrule
  \end{tabular}%
  }
\end{table*}

\rsection{2D-PT Parameter Variation with Instances and Noise}

The row ($\beta$) and column ($P$) parameters for 2D-PT are determined iteratively, beginning from an initial top-left replica where $\beta = \beta_0$ and $P = P_0$. For a given replica $(i,j)$ in the array, the standard deviations of its energy $\sigma_E$ and constraints $\sigma_g$ (evaluated by running short Markov chains) determine the parameters for the adjacent replicas, $\beta(i+1,j)$ and $P(i,j+1)$, as follows:
\begin{align}
\beta(i+1,j) &= \beta(i,j) + \alpha_\beta / \sigma_E \label{apt_2d_b} \\
P(i,j+1) &= P(i,j) + \alpha_P / (\beta(i,j) \sigma_g) \label{apt_2d_p} 
\end{align}
where $\alpha_\beta$ and $\alpha_P$ define the step size between consecutive replicas. This adaptive algorithm stops row expansion when $\sigma_E$ falls below a predefined threshold (set to one-tenth of the mean weight amplitude) and halts column expansion once all samples satisfy the constraints ($g=0$). The final $\beta$ and $P$ vectors are then derived by taking the median values of the arrays along the column and row directions, respectively. A detailed description of this algorithm is provided in Ref. \cite{delacour2025two}.

\begin{figure}[h]
    \centering
    \includegraphics[width=\linewidth]{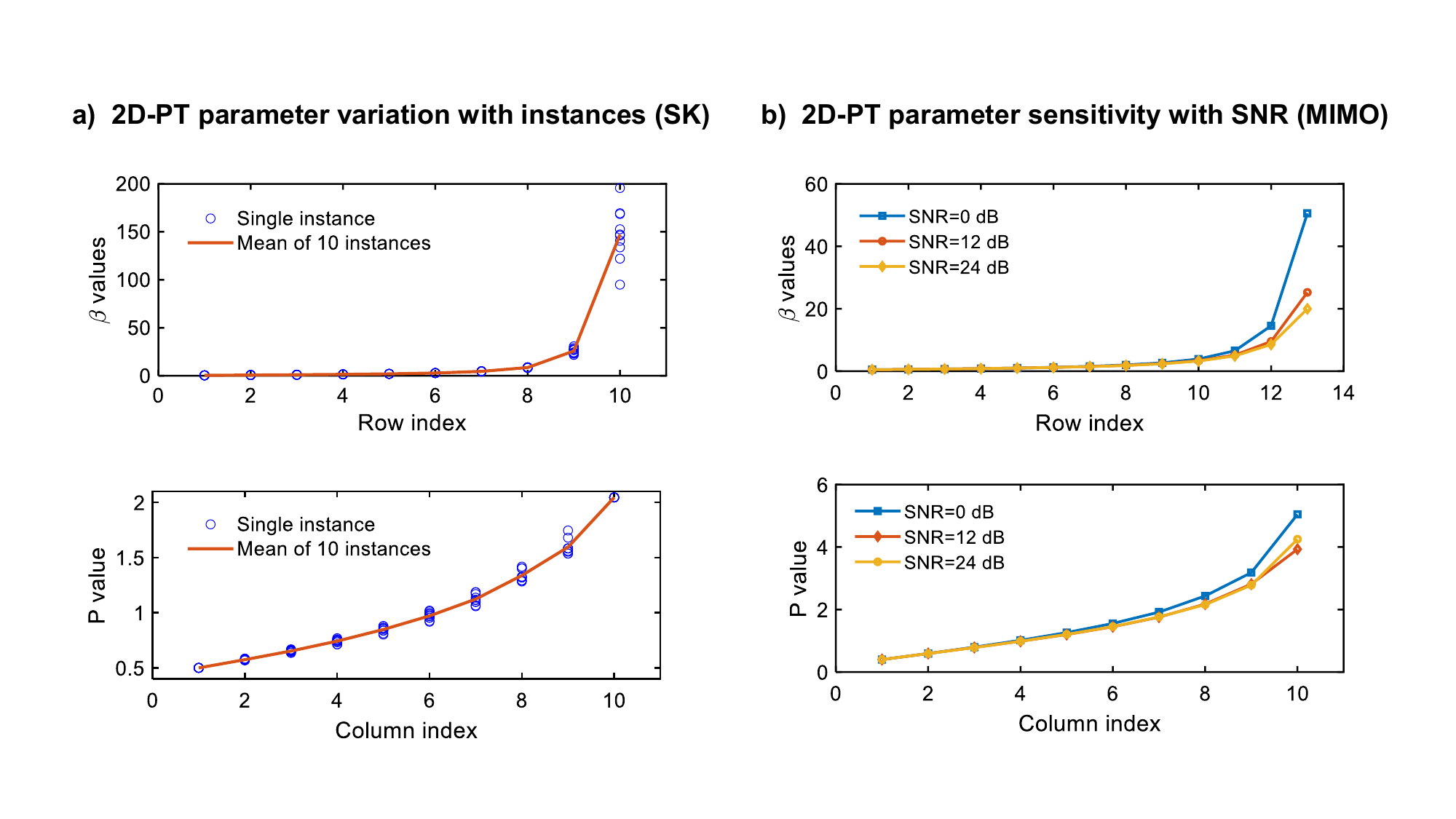}
    \caption{a) Variation of 2D-PT parameters ($\beta$,$P$) for 10 randomly generated Sherrington-Kirkpatrick (SK) instances ($N=64$). The resulting mean schedules are used throughout the SK experiment in the main FIG. 4. b) Sensitivity of 2D-PT parameters with signal-to-noise ratio (SNR) for MIMO instances ($N=64$). Each curve is obtained by averaging the 2D-PT schedules from 10 random MIMO instances at fixed SNR.}
    \label{fig_sup:2dpt_param_sensitivity}
\end{figure}

Figure \ref{fig_sup:2dpt_param_sensitivity}a illustrates the $\beta$ and $P$ values obtained for 10 random Sherrington-Kirkpatrick instances, sparsified using two copies per original spin (for a total of 128 spins). The adaptive procedure demonstrates robustness to variations in instances, except for the final row and column. These have significant fluctuations due to the inherently small standard deviations of the energy and constraints at large $\beta$ and $P$ values (as per Eq. \ref{apt_2d_b} and \ref{apt_2d_p}). Deriving new parameters for each instance is computationally expensive, requiring 20 independent chains of 1000 Monte Carlo sweeps (MCS) per replica here. To amortize this cost, we compute the mean parameters across 10 instances and apply them to the 100 unseen instances evaluated in the main manuscript.

Beyond instance variations caused by various channel matrices and transmitted symbols, the MIMO problem is also affected by noise corrupting the transmitted signals. As shown in Figure \ref{fig_sup:2dpt_param_sensitivity}b, the signal-to-noise ratio (SNR) directly impacts the adaptive algorithm's output for the 2D-PT parameters, with low SNR cases yielding larger parameter values. As before, to mitigate the computational overhead of the adaptive algorithm, the 2D-PT parameters used for MIMO throughout the manuscript are averaged over 10 random instances evaluated exclusively at SNR = 0.

\rsection{Iso-replica Comparison between 1D-PT and 2D-PT}
 
\begin{figure}[h]
    \centering
    \includegraphics[width=0.5\linewidth]{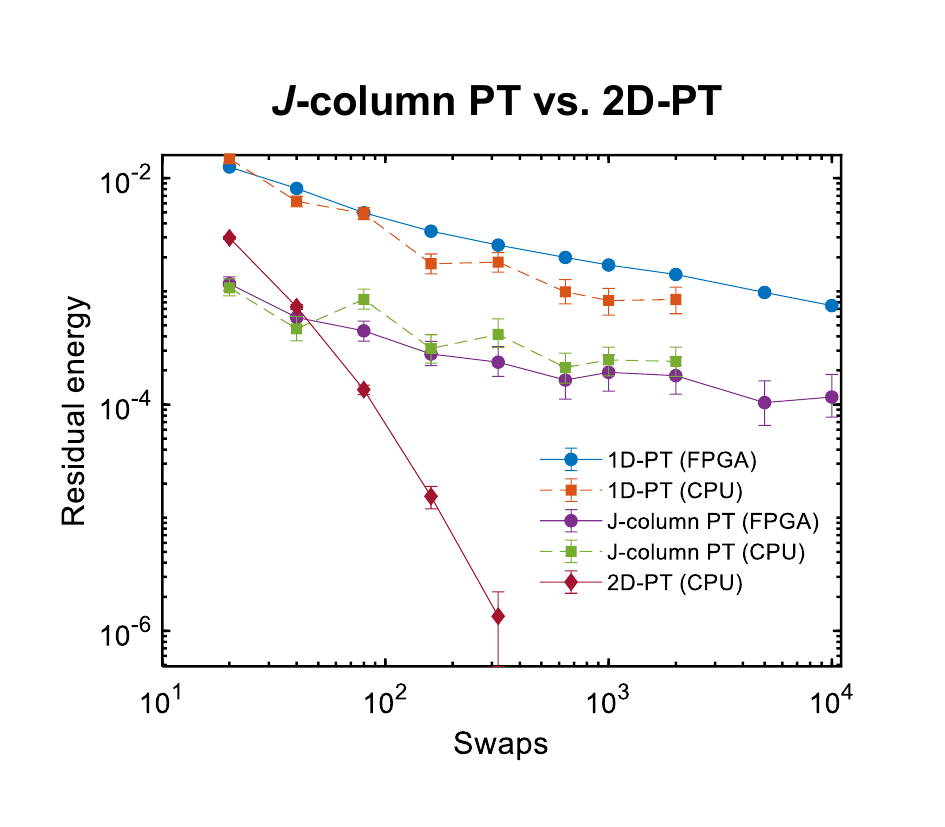}
    \caption{Iso-replica comparison between a parallelized 1D-PT ($10$ $\beta$ values, $J=10$ columns) and a 2D-PT approach, both having the same total number of replicas ($10\times 10$ array). The FPGA implementation has a fixed-point representation as detailed in Table~\ref{tab:params}, with residual energies exhibiting trends closely matching those of a double-precision (CPU) baseline. Data points are averaged across 100 instances and 100 runs per instance, with 95\% confidence intervals.}
    \label{fig_sup:2dpt_vs_Jcol}
\end{figure}

Compared to standard 1D-PT, 2D-PT introduces $J$ columns to explore increasing penalty strengths, $P_j$. Consequently, for a given set of $\beta$ parameters, 2D-PT uses $J$ times as many replicas as 1D-PT. A natural question is whether the performance advantage of 2D-PT stems from this increased replica count. We demonstrate that this is not the case through an iso-replica comparison. Using Sherrington-Kirkpatrick (SK) instances ($N=64$ spins, sparsified with two copies per spin), we execute a parallelized 1D-PT with $J=10$ columns to match the total number of replicas in the $10\times 10$ 2D-PT array. For each J-column PT data point in Figure~\ref{fig_sup:2dpt_vs_Jcol}, we apply the optimal penalty strength $P$ for the corresponding swap budget (derived from Fig.~3a in the main manuscript), and then record the minimum energy obtained across the 10 independent columns. Although parallelizing 1D-PT across independent columns lowers the residual energy by an offset, the energy reduction occurs at a significantly slower pace than in 2D-PT. Specifically, the $J$-column 1D-PT requires $10^4$ swaps to achieve a mean residual energy of $\rho_E^f \approx 10^{-4}$, whereas 2D-PT reaches this same value in roughly 80 swaps, yielding a $125\times$ speedup. Furthermore, 2D-PT successfully finds the constrained ground states in fewer than 400 swaps, a regime that remains completely inaccessible to the $J$-column approach even with a $10^4$ swap budget. These results confirm the fundamental algorithmic advantage of 2D-PT over parallel 1D-PT, consistent with findings previously reported in Ref.~\cite{delacour2025two} for sparsified Wishart instances.


\begin{thebibliography}{40}
\providecommand{\natexlab}[1]{#1}
\providecommand{\url}[1]{\texttt{#1}}
\expandafter\ifx\csname urlstyle\endcsname\relax
  \providecommand{\doi}[1]{doi: #1}\else
  \providecommand{\doi}{doi: \begingroup \urlstyle{rm}\Url}\fi

\bibitem[Chowdhury et~al.(2025)Chowdhury, Pieterse, Aadit, Mentink, and Camsari]{chowdhury2025probabilistic}
Shuvro Chowdhury, Jasper Pieterse, Navid~Anjum Aadit, Johan~H Mentink, and Kerem~Y Camsari.
\newblock Probabilistic computers for neural quantum states.
\newblock \emph{arXiv preprint arXiv:2512.24558}, 2025.

\bibitem[Mohseni et~al.(2022)Mohseni, McMahon, and Byrnes]{Mohseni2022}
Naeimeh Mohseni, Peter~L McMahon, and Tim Byrnes.
\newblock {Ising machines as hardware solvers of combinatorial optimization problems}.
\newblock \emph{Nature Reviews Physics}, 4\penalty0 (6):\penalty0 363--379, 2022.

\bibitem[Singh et~al.(2022)Singh, Jamieson, McMahon, and Venturelli]{singh2022ising}
Abhishek~Kumar Singh, Kyle Jamieson, Peter~L McMahon, and Davide Venturelli.
\newblock {Ising machines’ dynamics and regularization for near-optimal MIMO detection}.
\newblock \emph{IEEE Transactions on Wireless Communications}, 21\penalty0 (12):\penalty0 11080--11094, 2022.

\bibitem[Singh et~al.(2024)Singh, Kapelyan, Kim, Venturelli, McMahon, and Jamieson]{singh2024uplink}
Abhishek~Kumar Singh, Ari Kapelyan, Minsung Kim, Davide Venturelli, Peter~L McMahon, and Kyle Jamieson.
\newblock {Uplink MIMO detection using Ising machines: A multi-stage Ising approach}.
\newblock \emph{IEEE Transactions on Wireless Communications}, 2024.

\bibitem[Aadit et~al.(2022)Aadit, Grimaldi, Carpentieri, Theogarajan, Martinis, Finocchio, and Camsari]{aadit2022massively}
Navid~Anjum Aadit, Andrea Grimaldi, Mario Carpentieri, Luke Theogarajan, John~M Martinis, Giovanni Finocchio, and Kerem~Y Camsari.
\newblock Massively parallel probabilistic computing with sparse ising machines.
\newblock \emph{Nature Electronics}, 5\penalty0 (7):\penalty0 460--468, 2022.

\bibitem[Sajeeb et~al.(2025)Sajeeb, Aadit, Chowdhury, Wu, Smith, Chinmay, Raut, Camsari, Delacour, and Srimani]{sajeeb2025scalable}
M~Mahmudul~Hasan Sajeeb, Navid~Anjum Aadit, Shuvro Chowdhury, Tong Wu, Cesely Smith, Dhruv Chinmay, Atharva Raut, Kerem~Y Camsari, Corentin Delacour, and Tathagata Srimani.
\newblock {Scalable connectivity for Ising machines: Dense to sparse}.
\newblock \emph{Physical Review Applied}, 24\penalty0 (1):\penalty0 014005, 2025.

\bibitem[Delacour et~al.(2025)Delacour, Sajeeb, Hespanha, and Camsari]{delacour2025two}
Corentin Delacour, M~Mahmudul~Hasan Sajeeb, Jo{\~a}o~P Hespanha, and Kerem~Y Camsari.
\newblock {Two-dimensional parallel tempering for constrained optimization}.
\newblock \emph{Physical Review E}, 112\penalty0 (2):\penalty0 L023301, 2025.

\bibitem[Johnson et~al.(2011)Johnson, Amin, Gildert, Lanting, Hamze, Dickson, Harris, Berkley, Johansson, Bunyk, et~al.]{johnson2011quantum}
Mark~W Johnson, Mohammad~HS Amin, Suzanne Gildert, Trevor Lanting, Firas Hamze, Neil Dickson, Richard Harris, Andrew~J Berkley, Jan Johansson, Paul Bunyk, et~al.
\newblock {Quantum annealing with manufactured spins}.
\newblock \emph{Nature}, 473\penalty0 (7346):\penalty0 194--198, 2011.

\bibitem[King et~al.(2023)King, Raymond, Lanting, Harris, Zucca, Altomare, Berkley, Boothby, Ejtemaee, Enderud, et~al.]{king2023quantum}
Andrew~D King, Jack Raymond, Trevor Lanting, Richard Harris, Alex Zucca, Fabio Altomare, Andrew~J Berkley, Kelly Boothby, Sara Ejtemaee, Colin Enderud, et~al.
\newblock Quantum critical dynamics in a 5,000-qubit programmable spin glass.
\newblock \emph{Nature}, 617\penalty0 (7959):\penalty0 61--66, 2023.

\bibitem[McMahon et~al.(2016)McMahon, Marandi, Haribara, Hamerly, Langrock, Tamate, Inagaki, Takesue, Utsunomiya, Aihara, et~al.]{mcmahon2016fully}
Peter~L McMahon, Alireza Marandi, Yoshitaka Haribara, Ryan Hamerly, Carsten Langrock, Shuhei Tamate, Takahiro Inagaki, Hiroki Takesue, Shoko Utsunomiya, Kazuyuki Aihara, et~al.
\newblock {A fully programmable 100-spin coherent Ising machine with all-to-all connections}.
\newblock \emph{Science}, 354\penalty0 (6312):\penalty0 614--617, 2016.

\bibitem[Honjo et~al.(2021)Honjo, Sonobe, Inaba, Inagaki, Ikuta, Yamada, Kazama, Enbutsu, Umeki, Kasahara, Kawarabayashi, and Takesue]{honjo_2021}
Toshimori Honjo, Tomohiro Sonobe, Kensuke Inaba, Takahiro Inagaki, Takuya Ikuta, Yasuhiro Yamada, Takushi Kazama, Koji Enbutsu, Takeshi Umeki, Ryoichi Kasahara, Ken-ichi Kawarabayashi, and Hiroki Takesue.
\newblock 100,000-spin coherent {Ising} machine.
\newblock \emph{Science Advances}, 7\penalty0 (40):\penalty0 eabh0952, October 2021.
\newblock ISSN 2375-2548.

\bibitem[Fahimi et~al.(2021)Fahimi, Mahmoodi, Nili, Polishchuk, and Strukov]{fahimi_2021}
Z.~Fahimi, M.~R. Mahmoodi, H.~Nili, Valentin Polishchuk, and D.~B. Strukov.
\newblock Combinatorial optimization by weight annealing in memristive hopfield networks.
\newblock \emph{Scientific Reports}, 11\penalty0 (1):\penalty0 16383, August 2021.
\newblock ISSN 2045-2322.

\bibitem[Jiang et~al.(2023)Jiang, Shan, He, and Li]{jiang_2023}
Mingrui Jiang, Keyi Shan, Chengping He, and Can Li.
\newblock Efficient combinatorial optimization by quantum-inspired parallel annealing in analogue memristor crossbar.
\newblock \emph{Nature Communications}, 14\penalty0 (1):\penalty0 5927, September 2023.
\newblock ISSN 2041-1723.

\bibitem[He et~al.(2025)He, Hong, Ding, Lin, Lai, Fang, Gong, Hou, and Liang]{he2025hardware}
Yihan He, Ming-Chun Hong, Qiming Ding, Chih-Sheng Lin, Chih-Ming Lai, Chao Fang, Xiao Gong, Tuo-Hung Hou, and Gengchiau Liang.
\newblock A hardware demonstration of a universal programmable rram-based probabilistic computer for molecular docking.
\newblock \emph{Nature Communications}, 2025.

\bibitem[Mallick et~al.(2020)Mallick, Bashar, Truesdell, Calhoun, Joshi, and Shukla]{mallick2020using}
Antik Mallick, Mohammad~Khairul Bashar, Daniel~S Truesdell, Benton~H Calhoun, Siddharth Joshi, and Nikhil Shukla.
\newblock {Using synchronized oscillators to compute the maximum independent set}.
\newblock \emph{Nature communications}, 11\penalty0 (1):\penalty0 1--7, 2020.

\bibitem[Moy et~al.(2022)Moy, Ahmed, Chiu, Moy, Sapatnekar, and Kim]{moy20221}
William Moy, Ibrahim Ahmed, Po-wei Chiu, John Moy, Sachin~S Sapatnekar, and Chris~H Kim.
\newblock {A 1,968-node coupled ring oscillator circuit for combinatorial optimization problem solving}.
\newblock \emph{Nature Electronics}, 5\penalty0 (5):\penalty0 310--317, 2022.

\bibitem[Lo et~al.(2023)Lo, Moy, Yu, Sapatnekar, and Kim]{lo2023ising}
Hao Lo, William Moy, Hanzhao Yu, Sachin Sapatnekar, and Chris~H Kim.
\newblock An ising solver chip based on coupled ring oscillators with a 48-node all-to-all connected array architecture.
\newblock \emph{Nature Electronics}, 6\penalty0 (10):\penalty0 771--778, 2023.

\bibitem[Graber and Hofmann(2024)]{Graber_2024}
Markus Graber and Klaus Hofmann.
\newblock An integrated coupled oscillator network to solve optimization problems.
\newblock \emph{Communications Engineering}, 3\penalty0 (1):\penalty0 116, Aug 2024.
\newblock ISSN 2731-3395.

\bibitem[\vspace{0mm}Choi(2011)]{choi2011minor}
Vicky \vspace{0mm}Choi.
\newblock {Minor-embedding in adiabatic quantum computation: II. Minor-universal graph design}.
\newblock \emph{Quantum Information Processing}, 10\penalty0 (3):\penalty0 343--353, 2011.

\bibitem[Sugie et~al.(2020)Sugie, Yoshida, Mertig, Takemoto, Teramoto, Nakamura, Takigawa, Minato, Yamaoka, and Komatsuzaki]{sugie2020minor}
Yuya Sugie, Yuki Yoshida, Normann Mertig, Takashi Takemoto, Hiroshi Teramoto, Atsuyoshi Nakamura, Ichigaku Takigawa, Shin-ichi Minato, Masanao Yamaoka, and Tamiki Komatsuzaki.
\newblock {Minor-embedding heuristics for large-scale annealing processors with sparse hardware graphs of up to 102,400 nodes}.
\newblock \emph{arXiv preprint arXiv:2004.03819}, 2020.

\bibitem[Hamerly et~al.(2019)Hamerly, Inagaki, McMahon, Venturelli, Marandi, Onodera, Ng, Langrock, Inaba, Honjo, Enbutsu, Umeki, Kasahara, Utsunomiya, Kako, Kawarabayashi, Byer, Fejer, Mabuchi, Englund, Rieffel, Takesue, and Yamamoto]{hamerly_2019}
Ryan Hamerly, Takahiro Inagaki, Peter~L. McMahon, Davide Venturelli, Alireza Marandi, Tatsuhiro Onodera, Edwin Ng, Carsten Langrock, Kensuke Inaba, Toshimori Honjo, Koji Enbutsu, Takeshi Umeki, Ryoichi Kasahara, Shoko Utsunomiya, Satoshi Kako, Ken-ichi Kawarabayashi, Robert~L. Byer, Martin~M. Fejer, Hideo Mabuchi, Dirk Englund, Eleanor Rieffel, Hiroki Takesue, and Yoshihisa Yamamoto.
\newblock Experimental investigation of performance differences between coherent {Ising} machines and a quantum annealer.
\newblock \emph{Science Advances}, 5\penalty0 (5):\penalty0 eaau0823, May 2019.
\newblock ISSN 2375-2548.

\bibitem[Han and Li(2022)]{han2022low}
Ke~Han and Daokun Li.
\newblock Low-latency fpga design and implementation of hermitian matrix inversion based on partitioned systolic array for massive mimo.
\newblock In \emph{2022 IEEE International Conference on Integrated Circuits, Technologies and Applications (ICTA)}, pages 15--16. IEEE, 2022.

\bibitem[Chi et~al.(2025)Chi, Hsu, and Huang]{chi2025multilinear}
Jung-Chun Chi, Yi-Chieh Hsu, and Yuan-Hao Huang.
\newblock Multilinear generalized svd processor for multicast multiuser mimo systems.
\newblock \emph{IEEE Transactions on Circuits and Systems I: Regular Papers}, 2025.

\bibitem[Xiang et~al.(2025)Xiang, Liang, Wu, Hou, Wang, Meng, Wang, and Yang]{xiang2025area}
Siwei Xiang, Liyan Liang, Junfeng Wu, Jia Hou, Jiaxing Wang, Yishuo Meng, Jianfei Wang, and Chen Yang.
\newblock An area-efficient and reconfigurable accelerator for massive mimo systems.
\newblock \emph{IEEE Transactions on Very Large Scale Integration (VLSI) Systems}, 2025.

\bibitem[Casta{\~n}eda et~al.(2022)Casta{\~n}eda, Benini, and Studer]{castaneda2022283}
Oscar Casta{\~n}eda, Luca Benini, and Christoph Studer.
\newblock A 283 pj/b 240 mb/s floating-point baseband accelerator for massive mu-mimo in 22fdx.
\newblock In \emph{ESSCIRC 2022-IEEE 48th European Solid State Circuits Conference (ESSCIRC)}, pages 357--360. IEEE, 2022.

\bibitem[Sreedhara et~al.(2023)Sreedhara, Roychowdhury, Wabnig, and Srinath]{sreedhara2023mu}
Shreesha Sreedhara, Jaijeet Roychowdhury, Joachim Wabnig, and Pavan~Koteshwar Srinath.
\newblock {MU-MIMO Detection Using Oscillator Ising Machines}.
\newblock In \emph{2023 IEEE/ACM International Conference on Computer Aided Design (ICCAD)}, pages 1--9. IEEE, 2023.

\bibitem[Sagan and Roychowdhury(2022)]{sagan2022implementing}
Naomi Sagan and Jaijeet Roychowdhury.
\newblock Das: Implementing dense ising machines using sparse resistive networks.
\newblock In \emph{Proceedings of the 41st IEEE/ACM International Conference on Computer-Aided Design}, pages 1--9, 2022.

\bibitem[Krikidis(2024)]{krikidis2024mimo}
Ioannis Krikidis.
\newblock {MIMO with analogue 1-bit phase shifters: A quantum annealing perspective}.
\newblock \emph{IEEE Wireless Communications Letters}, 13\penalty0 (6):\penalty0 1571--1575, 2024.

\bibitem[Kim et~al.(2024)Kim, Singh, Venturelli, Kaewell, and Jamieson]{kim2024x}
Minsung Kim, Abhishek~Kumar Singh, Davide Venturelli, John Kaewell, and Kyle Jamieson.
\newblock {X-ResQ: Reverse annealing for quantum MIMO detection with flexible parallelism}.
\newblock \emph{arXiv preprint arXiv:2402.18778}, 2024.

\bibitem[Zhu et~al.(2026)Zhu, Singh, Laydevant, Wu, Kapelyan, Venturelli, Jamieson, and McMahon]{Roumin2026fully}
Ruomin Zhu, Abhishek~Kumar Singh, J{\'e}r{\'e}mie Laydevant, Fan~O Wu, Ari Kapelyan, Davide Venturelli, Kyle Jamieson, and Peter~L McMahon.
\newblock A fully parallel densely connected probabilistic ising machine with inertia for real-time applications.
\newblock \emph{arXiv preprint arXiv:2604.17109}, 2026.

\bibitem[Delacour(2025)]{delacour2025self}
Corentin Delacour.
\newblock Self-adaptive ising machines for constrained optimization.
\newblock In \emph{2025 Design, Automation \& Test in Europe Conference (DATE)}, pages 1--7. IEEE, 2025.

\bibitem[Camsari et~al.(2017)Camsari, Faria, Sutton, and Datta]{camsari2017stochastic}
Kerem~Yunus Camsari, Rafatul Faria, Brian~M Sutton, and Supriyo Datta.
\newblock {Stochastic p-bits for invertible logic}.
\newblock \emph{Physical Review X}, 7\penalty0 (3):\penalty0 031014, 2017.

\bibitem[Poor(2000)]{poor2000turbo}
H~Vincent Poor.
\newblock Turbo multiuser detection: An overview.
\newblock In \emph{2000 IEEE Sixth International Symposium on Spread Spectrum Techniques and Applications. ISSTA 2000. Proceedings (Cat. No. 00TH8536)}, volume~2, pages 583--587. IEEE, 2000.

\bibitem[Verdu(1998)]{verdu1998multiuser}
Sergio Verdu.
\newblock \emph{Multiuser detection}.
\newblock Cambridge university press, 1998.

\bibitem[Swendsen and Wang(1986)]{swendsen1986replica}
Robert~H Swendsen and Jian-Sheng Wang.
\newblock Replica monte carlo simulation of spin glasses.
\newblock \emph{Physical review letters}, 57\penalty0 (21):\penalty0 2607--2609, 1986.

\bibitem[Hukushima and Nemoto(1996)]{hukushima1996exchange}
Koji Hukushima and Koji Nemoto.
\newblock Exchange monte carlo method and application to spin glass simulations.
\newblock \emph{Journal of the Physical Society of Japan}, 65\penalty0 (6):\penalty0 1604--1608, 1996.

\bibitem[Aadit et~al.(2023)Aadit, Mohseni, and Camsari]{aadit2023accelerating}
Navid~Anjum Aadit, Masoud Mohseni, and Kerem~Y Camsari.
\newblock {Accelerating Adaptive Parallel Tempering with FPGA-based p-bits}.
\newblock In \emph{2023 IEEE Symposium on VLSI Technology and Circuits (VLSI Technology and Circuits)}, pages 1--2. IEEE, 2023.

\bibitem[Nikhar et~al.(2024)Nikhar, Kannan, Aadit, Chowdhury, and Camsari]{nikhar2024all}
Srijan Nikhar, Sidharth Kannan, Navid~Anjum Aadit, Shuvro Chowdhury, and Kerem~Y Camsari.
\newblock All-to-all reconfigurability with sparse and higher-order ising machines.
\newblock \emph{Nature Communications}, 15\penalty0 (1):\penalty0 8977, 2024.

\bibitem[Clark et~al.(2016)Clark, Vashishtha, Shifren, Gujja, Sinha, Cline, Ramamurthy, and Yeric]{clark2016asap7}
Lawrence~T Clark, Vinay Vashishtha, Lucian Shifren, Aditya Gujja, Saurabh Sinha, Brian Cline, Chandarasekaran Ramamurthy, and Greg Yeric.
\newblock Asap7: A 7-nm finfet predictive process design kit.
\newblock \emph{Microelectronics Journal}, 53:\penalty0 105--115, 2016.

\bibitem[Carsello et~al.(2022)Carsello, Thomas, Nayak, Chen, Horowitz, Raina, and Torng]{carsello2022mflowgen}
Alex Carsello, James Thomas, Ankita Nayak, Po-Han Chen, Mark Horowitz, Priyanka Raina, and Christopher Torng.
\newblock mflowgen: A modular flow generator and ecosystem for community-driven physical design.
\newblock In \emph{Proceedings of the 59th ACM/IEEE Design Automation Conference}, pages 1339--1342, 2022.

\end{thebibliography}
\end{document}